\documentclass[10pt,journal,compsoc]{IEEEtran}
\IEEEoverridecommandlockouts

\usepackage{cite}
\usepackage{amsmath,amssymb,amsfonts,hyperref}
\usepackage{algorithmic}
\usepackage{graphicx}
\usepackage{textcomp}
\usepackage{comment}
\usepackage{blindtext}
\usepackage{booktabs} 
\usepackage{multirow}
\usepackage{graphics}
\usepackage{subfigure}
\usepackage{tikz}
\usepackage{enumitem}
\usepackage{xspace}
\usepackage[utf8]{inputenc}
\usepackage{makecell}
\usepackage{balance}
\usepackage{wrapfig,lipsum}
\usepackage{color}
\usepackage{colortbl}
\usepackage{listings}
\usepackage{url}
\usepackage{mathenv}
\usepackage{multicol}
\usepackage{setspace}
\usepackage{mathtools}

\newcommand{\etal}{{\em et al.}}

\newcommand{\ie}{{\em i.e.,}\xspace}

\newcommand{\BfPara}[1]{\vspace{1mm}{\noindent\bf#1.}\xspace}
\newcommand*\cib[1]{\tikz[baseline=(char.base)]{
                            \node[shape=circle,color=black, fill=black!70!white,text=white,draw,inner sep=0.3pt] (char) {#1};}}
\newcommand{\citet}[1]{\citeauthor{#1} \shortcite{#1}}

\newcommand{\tma}{{\sf TM-1}\xspace}
\newcommand{\tmb}{{\sf TM-2}\xspace}
\newcommand{\tmc}{{\sf TM-3}\xspace}
\newcommand{\exploreSegments}{ \mbox{$\textsc{ExploreSegments()}$ }}

\definecolor{brightmaroon}{rgb}{0.76, 0.13, 0.28}
\definecolor{darkgreen}{rgb}{0.0, 0.2, 0.13}

   \hypersetup{
	colorlinks,
	urlcolor=blue,
	linkcolor=red,
	citecolor=green,
	bookmarksnumbered
}

\usepackage{xcolor}

\begin{document}
\title{Learning Location from Shared Elevation Profiles in Fitness Apps: A Privacy Perspective}

\author{Ulku Meteriz-Yildiran, Necip Fazil Yildiran, Joongheon Kim, and David Mohaisen
\IEEEcompsocitemizethanks{\IEEEcompsocthanksitem U. Meteriz-Yildiran is with Meta and N. Yildiran is with Google; the work of both authors was done while they were at the University of Central Florida. D. Mohaisen is with the Department of Computer Science, University of Central Florida, Orlando, FL 32816, USA. J. Kim is with the Department of Electrical Engineering at Korea University, Republic of Korea.  D. Mohaisen is the corresponding author (e-mail: mohaisen@ucf.edu). 
An earlier version of this work has appeared in IEEE ICDCS  2020~\cite{MeterizYKM20}. This work was supported in part by NRF under grant 2016K1A1A2912757 and CyberFlorida Seed Grant (2021/2022). J. Kim was supported by NRF under grant 2022R1A2C20048690. J. Kim (joongheon@korea.ac.kr) and D. Mohaisen (mohaisen@ucf.edu) are the corresponding authors. 
}
}




\IEEEtitleabstractindextext{%
\begin{abstract}
The extensive use of smartphones and wearable devices has facilitated many useful applications. 
For example, with Global Positioning System (GPS)-equipped smart and wearable devices, many applications can gather, process, and share rich metadata, such as geolocation, trajectories, elevation, and time. 
For example, fitness applications, such as Runkeeper and Strava, utilize the information for activity tracking and have recently witnessed a boom in popularity. 
Those fitness tracker applications have their own web platforms and allow users to share activities on such platforms or even with other social network platforms. 
To preserve the privacy of users while allowing sharing, several of those platforms may allow users to disclose partial information, such as the elevation profile for an activity, which supposedly would not leak the location of the users. 
In this work, and as a cautionary tale, we create a proof of concept where we examine the extent to which elevation profiles can be used to predict the location of users. 
To tackle this problem, we devise three plausible threat settings under which the city or borough of the targets can be predicted. 
Those threat settings define the amount of information available to the adversary to launch the prediction attacks. Establishing that simple features of elevation profiles, e.g., spectral features, are insufficient, we devise both natural language processing (NLP)-inspired text-like representation and computer vision-inspired image-like representation of elevation profiles, and we convert the problem at hand into text and image classification problem. 
We use both traditional machine learning- and deep learning-based techniques and achieve a prediction success rate ranging from 59.59\% to 99.80\%. 
The findings are alarming, highlighting that sharing elevation information may have significant location privacy risks.
  
\end{abstract}
}
\maketitle

\section{Introduction}

From smartphones to wearables, an increasing number of Internet of Things (IoT) devices are equipped with Global Positioning System (GPS), accelerometers, and gyroscopes to allow applications to function or to present a better user experience using {\em geodata}, such as location and elevation information.
More recently, fitness applications that run on smartphones and smartwatches used these components to collect spatial, temporal, and activity-specific information to analyze, summarize, and visualize users' activities. 
By analyzing each activity, many of those applications  deliver personalized motivations and challenges for users to meet their goals. 
Using social media support of these applications for sharing updates about users' activities, including training routes and elevation profiles for the routes taken for an activity (e.g., walking, running, climbing, cycling), users can have positive behavioral changes through a more active lifestyle motivated by competitions with acquaintances~\cite{higgins}.

Despite the broad set of advantages that geodata offers, geodata usage and uncontrolled sharing can pose a significant privacy risk that can be further exploited in multiple attacks, including stalking \cite{Polakis:2015:WWP:2810103.2813605} and cybercasing \cite{Cybercasing}. 
For example, with a large amount of geotagged data, including text, images, and videos, cybercasing provides criminals and maliciously motivated individuals with a significant attack vector. 
Geo-tagged photos that are frequently posted on image-sharing websites, such as Flickr, or second-hand sale websites, such as Craigslist, may put owners of those images at risk. 
For example, geo-tagged images posted on sales websites may reveal the location of the advertised product, leading to trespassing or even theft.


\begin{figure}[t]
    \centering
    \includegraphics[width=0.45\textwidth]{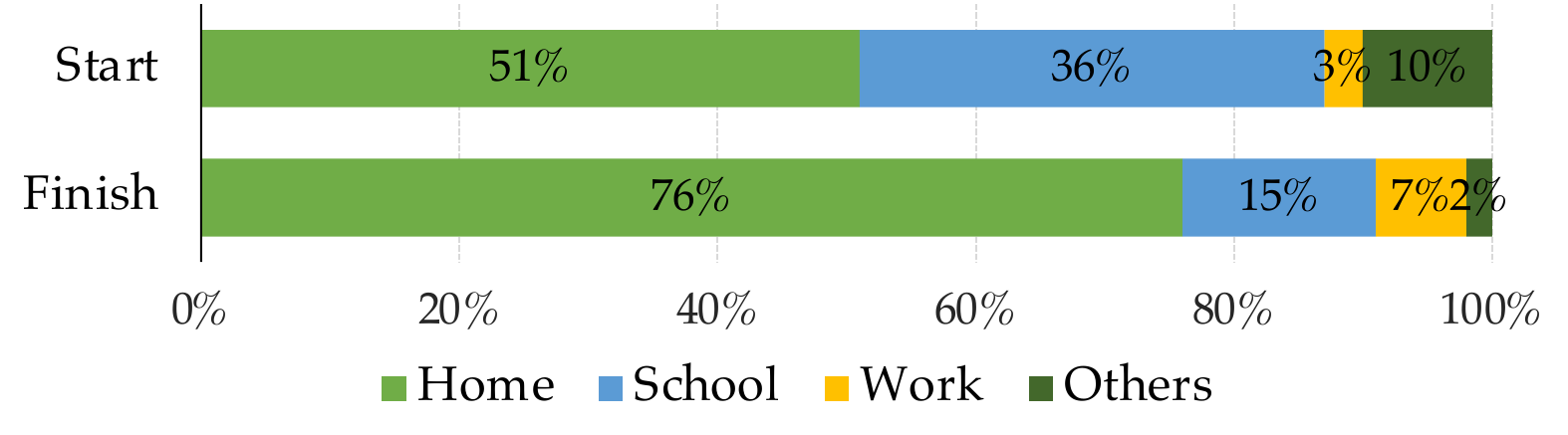}\vspace{-3mm}
    \caption{Survey results for understanding users' behavior with starting point statistics and finishing point statistics. While $90\%$ of the $60$ participants indicated their start of activity is either home, school, or work, an overwhelming $98\%$ of the participant indicated those to be the finish (end) point of their activities.}
    \label{figure:survey}\vspace{-5mm}
\end{figure}

While geodata recorded by fitness applications is indeed important and valuable for the operation of those applications, this data can also be used for launching attacks on users by breaching their privacy since sensitive information of users, such as home or workplace location, can be easily inferred from such data. 
Even worse, a large number of users, when sharing such information, would be unaware of the ramifications of sharing and the potential risk of inferring such contextual information, such as home, work location, etc., from such shared location data. 
To support this argument, we conducted an online survey with $60$ participants who regularly use fitness applications outdoors.
The results of the survey, summarized in \autoref{figure:survey}, reveal that $51\%$ of the participants start their training from their homes, $36\%$ start from their school, and $3\%$ start from their workplace, while $76\%$ of the participants finish their training at their homes. 
Moreover, for the same set of users (results are not shown in \autoref{figure:survey}), $42\%$ of those users have indicated that not sharing location information implies privacy protection, while $30\%$ of the respondent were uncertain, and $28\%$ were certain that not sharing would not necessarily mean their privacy is protected. 
The mixed responses highlight the gap between reality and expectations of privacy when sharing location information online and call for further investigation. 

Although it is possible to hide the location trajectory by removing the activity map in the fitness applications, users still want to share elevation profiles or certain statistics of the activity to show the roughness, technicality, and difficulty of the routes they took as a measure of their workout. 
For example, up until recently, users have been demanding those fitness applications to allow for fine-grained and customized access control by allowing them to share the elevation profile of an activity while masking the map that highlights the actual trajectory, which is deemed of high privacy value to them~\cite{forum1, forum2, forum3, forum4}.

In the same survey we conducted earlier, we asked our $60$ subjects ``while sharing an outdoor workout record, do you think hiding the map and sharing only the statistics of your training (such as speed and elevation changes) is enough for protecting your privacy?''. 
The results were overwhelmingly positive, with $25$ of them indicating ``yes'', $18$ indicating ``maybe'' (together accounting for more than $71\%$), and only $17$ indicating ``no''. 

Is sharing the elevation profile of activity enough to maintain the privacy of users? In this paper, we argue that an approximate location, extracted from the contexts of activities and at different levels of location granularity, could still be revealed from the elevation profile information. 
We examine this problem comprehensively and develop techniques that can be used to accurately associate an elevation profile with contextual information, such as the location.

\BfPara{Contributions} In this paper, we contribute the following: 

\begin{itemize}[leftmargin=*]
    \item we translate the problem of location privacy inference from elevation profiles into text classification and image classification problems by encoding the elevation signals as strings and visualizing the elevation signals as images to employ various common approaches for solving image and text classification problems, 
    \item we investigate the possible attack surface for the problem by exploring three different threat models, which we later use to evaluate the success of our approaches by simulating our methods considering each threat model, 
    \item we demonstrate that location information can be predicted from elevation profile using different machine/deep learning methods with accuracy in the range $80.25\% - 99.80\%$ at different resolutions.
\end{itemize}

We note that examining the effect of the attack using a large-scale in-the-wild case study is impractical as service providers prevent the use of their data for tracking by a third party. However, to motivate the effect of the attack, we consider the scenario of an informed adversary who knows the city where a victim with the exposed elevation profiles for associated activities lives. As such, the adversary proceeds by profiling the city and collecting elevation profiles for different segments within the city. One can see how easily such an adversary will be able to contextualize the elevation profiles of the victim’s activity further by narrowing it down to a few candidate precomputed elevation profiles. Given the adversary’s awareness of the mapping between the location and the profiles, the adversary will be able to easily infer valuable information about the habits of the victim by associating, for instance, end, start, and stopping points on the elevation profile, with points of interest (cafes, workplace, etc.). 


\BfPara{Organization} We present the background in \autoref{sec:background}, the threat model in \autoref{sec:threat}, a high-level overview of our approach in \autoref{sec:hlapproach}, the implementation details are presented in \autoref{sec:implementation}, the evaluation results in \autoref{sec:results}, further discussions in \autoref{sec:discussion}, the related work in \autoref{sec:related}, and concluding remarks in \autoref{sec:conclusion}.

\section{Background}\label{sec:background} 

In this section, we provide some background information highlighting the significance of elevation profiles for athletes, the use cases, some properties of the fitness applications on the market today, and some reported privacy breach incidences of fitness applications to contextualize further the work presented in the rest of this paper. 

\subsection{Elevation Profiles Importance for Athletes}
Athletes who keep track of their activity records measure various modalities and attributes associated with the activities, including the distance, speed, overall time, and heart rate over the course of the activity. 
Based on these attributes, they adjust their training strategies to reach their goals. 
Elevation changes, often reported in the form of elevation gain, are one of the most significant attributes measuring the performance of a cyclist/runner and often depict how hard the run or ride is.
For example, riding a bike for a 20-mile ride while climbing 1000 feet in total is significantly more challenging than biking on a flat terrain~\cite{Padilla99}.
Therefore, when recording or sharing a ride/run, athletes care about the changes in the elevation, thus elevation profiles.


\begin{table}[t]
    \caption{Popular fitness applications and their features. {ET}: Exercise tracking. 
    {SS}: Ability to share to social media. 
    {SNS}: Social networking capabilities in the service.
    {PR}: Private records. 
    {BU}: User blocking capability.}
    \centering
    \begin{tabular}{|c|c|c|c|c|c|}
        \hline
        Service & ET & \makecell{SS} & SNS & PR & BU \\
        \hline
        Strava         & $\bullet$ & $\bullet$ & $\bullet$ & $\bullet$ & $\bullet$ \\
        \hline
        Runtastic      & $\bullet$ & $\bullet$ & $\bullet$ &     $\circ$       & $\bullet$ \\
        \hline
        Runkeeper      & $\bullet$ & $\bullet$ & $\bullet$ & $\bullet$ &$\circ$ \\
        \hline
        Nike+ Running  & $\bullet$ & $\bullet$ & $\bullet$ & $\bullet$ &$\circ$ \\
        \hline
        MapMyRun       & $\bullet$ & $\bullet$ & $\bullet$ & $\bullet$ &$\circ$ \\
        \hline
    \end{tabular}
    \label{table:fitness_applications}
\end{table}


\subsection{Fitness Applications \& Privacy Breach Incidents}

Fitness applications allow users to track their workout history and provide them with statistics.
Moreover, some fitness applications have social network capabilities, as shown in \autoref{table:fitness_applications}, and allow users to share workout summaries that are known to motivate users and their social network connections to achieve their goals~\cite{higgins}. 
Some fitness applications also inherit user-blocking features and capabilities from social network platforms, including user privacy options such as private records--the activity records that are only visible to the user.

Although fitness applications have configurable privacy options, there have been a lot of privacy incidents concerning location data obtained from those fitness applications. 
We review some of those privacy breaches in the following to contextualize our work in the broader privacy literature. 

\BfPara{Revealing Secret U.S. Military Bases} Strava, which is one the most popular fitness tracking applications in the market today, collects users' public data and publishes a heatmap of the aggregates to highlight routes frequented by users~\cite{StravaHeatmap}. 
Although the aggregates in the heatmap do not explicitly contain any identity information, activities in desolate places revealed the location of many U.S. military bases, which is considered sensitive information~\cite{StravaMilitaryBaseIncidentWashingtonPost,StravaMilitaryBaseIncidentTheGuardian}.

\BfPara{Deanonymization Through Strava Segments}
In Strava, the heatmap feature was used to show ``heat'' made by the aggregated and public activities of Strava users over the past year. It is, however, shown that a dedicated adversary can deanonymize heatmap to find out users who ran in a specified route~ \cite{AdvancedDeanonymizationThroughStrava}. 
For example, by selecting a route from the heatmap, a registered user can manually create a GPS eXchange (GPX) track file and create a segment using it on Strava.
A segment is a portion of a road or a trail where athletes compare their finishing times.
Consequently, once this segment is created, the users who previously ran that route are shown on the leaderboard grouped by gender and age.
This feature is then leveraged to identify individuals who ran that particular place.

\BfPara{Tracking and Bicycle Theft} 
Users of fitness applications can share information related to the equipment used for the activity, including bicycles, tracking devices, shoes, etc., along with the routes frequented. The combined shared information makes them a target for robbery, and several such incidents of bicycle theft are reported~\cite{bike1,bike2,bike4,bike5}.

\BfPara{Attack on Privacy Zone}
To cope with the increasing privacy risks, Strava features {\em privacy zones}, a technique to obfuscate the exact start and end points of a route. 
A recent study~\cite{217618} has demonstrated that it is possible to reveal the exact start and end point of a route that utilizes the privacy zone feature. 
The same study also claimed that around $95\%$ of the users are at risk of revealing their location information.

\BfPara{Live Activity Breach}
In Runtastic, one of the popular activity-tracking applications, users can share their live activities.
In theory, users should be able to configure the privacy settings for their activities such that only privileged users, such as connections on the application platform, can track the shared live activity session.
However, it has been demonstrated~\cite{Runtastic_privacy} that the selected privacy settings are not correctly applied to a live session.
As a result, everyone can go through live sessions and track Runtastic users in real time, even though the associated privacy options should have prevented this type of breach.
Based on this incident, it would be easy to stalk and locate a user, e.g., a lone runner or cyclist with expensive equipment, in real time.

%
\section{Threat Models}\label{sec:threat}
We outline the potential threat models under which this study is conducted. 
We describe three models under which location privacy is breached only from associated elevation profiles.
We note that the following threat models are only hypothetical: no attacks were actually launched on any users. 
As mentioned earlier, this study in its entirety is motivated by the aforementioned demands of users to have more flexibility over-sharing partial data, such as elevation profiles, and examines the ramifications of such sharing in a hypothetical setting. 
We note, however, that those settings are also plausible if such sharing is enabled. 

Our study utilizes three threat models: \tma, \tmb, and \tmc, which we outline below with their justifications. The adversarial capabilities in \tma are greater than in \tmb and \tmc, making it a more restrictive (powerful) model. 

\BfPara{\cib{1} \tma} In \tma, we assume an adversary with workout history records of a target user, and the goal of the adversary is to identify the last workout location of the target user from the recently shared elevation profiles. 
\tma is justified by multiple plausible scenarios in practice. 
For example, such an adversary might have been a previous social network connection of the target user that was later blocked. 
In such a scenario, the adversary may have previous workout records of the target from which the adversary may attempt to de-anonymize the target's activities.
Another example might include group activities, where two individuals (i.e., the adversary and target) may have shared the same route at some point. 
In either case, by knowing the target's previous fitness activity records, the main goal of the adversary in this model is to identify recent whereabouts only from publicly shared elevation profiles in workout summaries, thus breaching the target's location privacy. 
 
\BfPara{\cib{2} \tmb} In \tmb, we assume an adversary with access to limited information, such as the city where the target lives. 
Such information is easily accessible from public profile summaries, \url{athlinks.com}, public records, etc. 
The adversary's goal in \tmb is to find out which region or part of a given city the target's activities are associated with. 
The \tmb use scenario may include a targeted user sharing private activities in which the route is hidden while the elevation profile is shown. 
The adversary, knowing the city where the target lives, would want to identify the region (e.g., a borough in the city) associated with the user's activity.

\BfPara{\cib{3} \tmc} In \tmc, we assume an adversary trying to identify the target user's city using only publicly shared elevation profiles without any prior information.
We assume, however, the adversary has the ability to profile the elevation of cities with information that is easily obtained from public sources (e.g., Google Maps, OpenStreetMap). 
The use scenario of \tmc may be used as a stepping stone towards launching the attack scenario in \tmb upon narrowing down the search space to a city.

\section{Approach: High-Level Overview}\label{sec:hlapproach}

\begin{figure*}[t]
    \centering
    \includegraphics[width=1.0\textwidth]{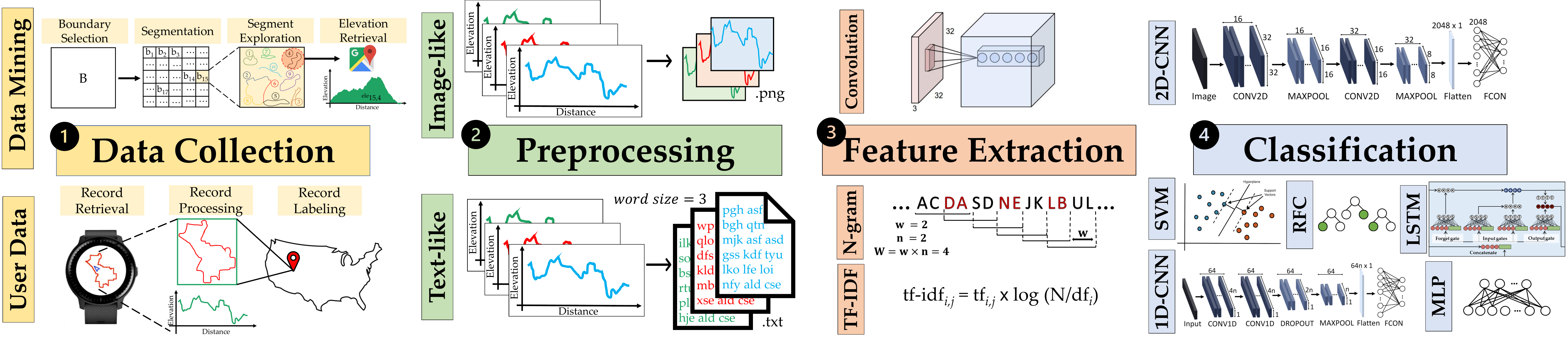}
    \caption{The end-to-end pipeline of the approach consists of four main stages: (i) data collection, (ii) preprocessing, (iii) feature extraction, and (iv) classification. There are two types of data representations, each of which is processed differently in the feature extraction and classification stages. The feature extraction of the image-like representation is internally handled in the convolution layers of the classification phase. The convolution illustration in the feature extraction step is for the sake of modularization and for consistency with the other pipeline instance.}
    \label{figure:overall_pipeline}
\end{figure*}

In this section, we give a brief overview of our pipeline, which consists of the data collection, preprocessing, feature extraction, and classification as illustrated in \autoref{figure:overall_pipeline}.
Each phase of the pipeline is detailed in \autoref{sec:implementation}.

\BfPara{Data Collection} 
We collected three datasets with varying and rich characteristics, namely (i) user-specific activity data collected from an athlete, (ii) mined training route segments grouped at city-level, and (iii) mined training route segments grouped at borough-level. 
For the user-specific dataset, we collected physical activity records of athletes and converted those activities to an intermediate format, the GPS Exchange Format (GPX). 
Then, we parsed the GPX files and manually labeled them according to the latitude and longitude information included within each file. 
For the second dataset, we mined training route segments from a popular fitness tracking website by specifying the location boundaries, \ie the class label of the mined data, and augmented each segment with the corresponding elevation profiles obtained from Google Maps Elevation API. 
Finally, we similarly constructed the borough-level dataset as in the city-level dataset.

\BfPara{Preprocessing} 
We employ Natural Language Processing (NLP) and computer vision techniques to convert the problem to text classification and image classification problems, respectively.
To this end, we prepare the data accordingly in the preprocessing phase.
Preprocessing consists of two parts: (i) text-like and (ii) image-like representations. 

For text-like representation, we discretize the elevation signals and compute the minimum required \emph{word} size. 
We then create a mapping between each unique discrete value and a string. 
By mapping the string correspondents to the unique discrete values, we encode the elevation profiles in text. 
We, then, form a \emph{vocabulary} from the text sequences of each dataset using the $n$-grams. 

To obtain image-like representations, we convert the elevation profiles to a fixed-sized line graph where the \textit{x}-axis stands for time and the \textit{y}-axis stands for the elevation values. 
We also color the lines in the graphs to represent the elevation interval in which the elevation profiles range.

\BfPara{Feature Extraction} 
The classification algorithms operate on high-quality and discriminative features obtained from the representations of elevation profiles. 
For feature extraction, we utilize NLP and computer vision approaches. 

To employ NLP approaches using the vocabulary obtained in preprocessing phase, we represent each elevation profile as either a feature vector based on the vocabulary frequency in the text-like representation (bag-of-words vector) or as a term frequency-inverse document frequency (tf-idf) vector. 
To employ computer vision approaches, we utilize Convolutional Neural Networks (CNN) over image-like representations. 
The optimal features of an image-like representation are efficiently extracted by the convolutional and pooling layers in the CNN architecture.

\BfPara{Multi-Class Classification} 
We utilize various machine learning and deep learning models for classification, including Support Vector Machine (SVM) and Random Forest Classification (RF), Multi-Layer Perceptron (MLP), Long Short-Term Memory (LSTM), 1D Convolutional Neural Network (C1D), and 2D Convolutional Neural Network (CNN).

\section{Implementation Details}\label{sec:implementation}
The implementation details of data collection, preprocessing, feature extraction, and multi-class classification are addressed in the following subsections. 

\subsection{Data Collection}
In this study, we compiled three datasets: the user-specific dataset, the city-level dataset, and the borough-level dataset. 
The user-specific dataset is retrieved from a voluntary athlete who frequently records activities through fitness applications. 
It offers dense and thorough coverage of regions frequented by the user; those regions are used as class labels. 
The city-level and borough-level datasets are created from scratch by collecting location trajectories that are created and frequented by the athletes. 
Both city-level and borough-level datasets provide sparse coverage of cities and boroughs.

\subsubsection{User-Specific Dataset}
\begin{figure}[t]
\centering
\includegraphics[width=0.4\textwidth]{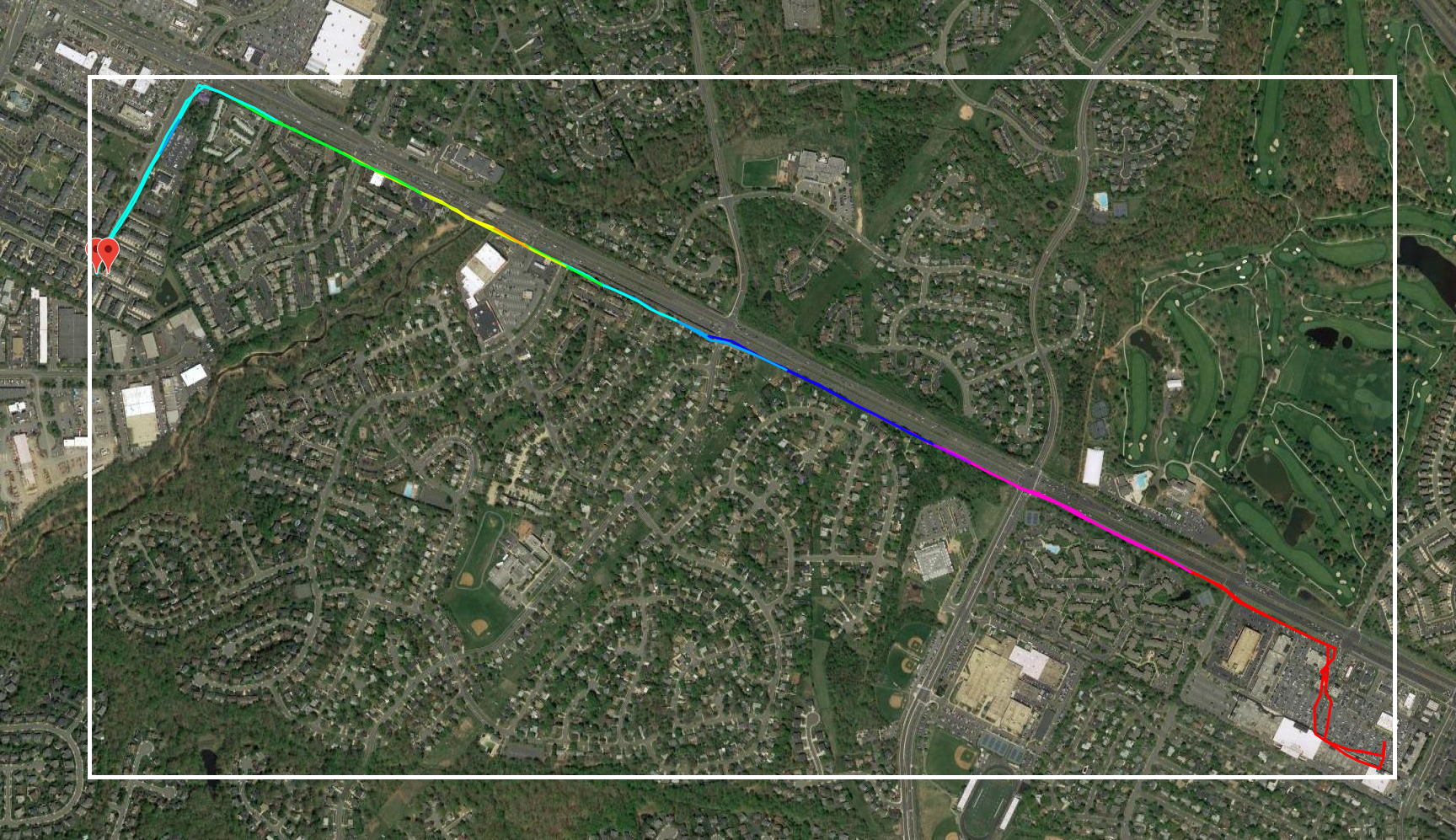}
\caption{An illustration of the tight rectangle encapsulating an example route. The rectangle is created by fitting the route in between the minimum and maximum $(latitude, longitude)$ pairs of the given route. The minimum and maximum $(latitude, longitude)$ pairs correspond to the bottom left (\ie South West) corner and the top right (\ie North East) corner of the rectangle, respectively.}
\label{figure:route}
\end{figure}

For the user-specific dataset, we collected activity data, including each activity's location trajectory and the corresponding elevation profile from a voluntary athlete who records activities frequently through fitness applications. 
First, the location trajectories included in the user-specific dataset are converted to GPX format to avoid confusion caused by different formats and settings across the activity records. 
Then, to label the samples, the maximum and minimum coordinates of each location trajectory are fetched. 
Each sample location trajectory is encapsulated with a tight rectangle whose top right (North East) and bottom left (South West) corners are computed from the maximum and minimum coordinates of the trajectory as illustrated in  \autoref{figure:route}. 
To classify the samples, each rectangle encapsulating the trajectory is compared with the previously created regions. 
If the Euclidean distance between the center of the rectangle and the center of the existing region does not exceed a predetermined threshold, the rectangle and its corresponding sample are labeled with a unique identity of the region. Then, we annotated the region labels, such as Orlando, Washington DC etc., based on the manual observation on the map. 
If no region includes the trajectory, a new region is created.
The final sample size distribution of the user-specific dataset is shown in \autoref{table:raw}. 
    
The user-specific dataset is prone to have similar location trajectory portions across its samples since the user may frequent the same set of places in his/her everyday activities, such as the location trace they follow while leaving/arriving home or their favorite routes. 
Therefore, we calculated the average overlap ratio of the routes included in the user-specific dataset by comparing each sample with the other samples with the same class label. 
For each sample pair comparison, the overlap ratio is calculated as the intersection-over-union of the tight rectangles encapsulating the sample routes. 
The average overlap ratio of the user-specific dataset is calculated as $35\%$.

\begin{figure}[t]
\centering
    \includegraphics[width=0.48\textwidth]{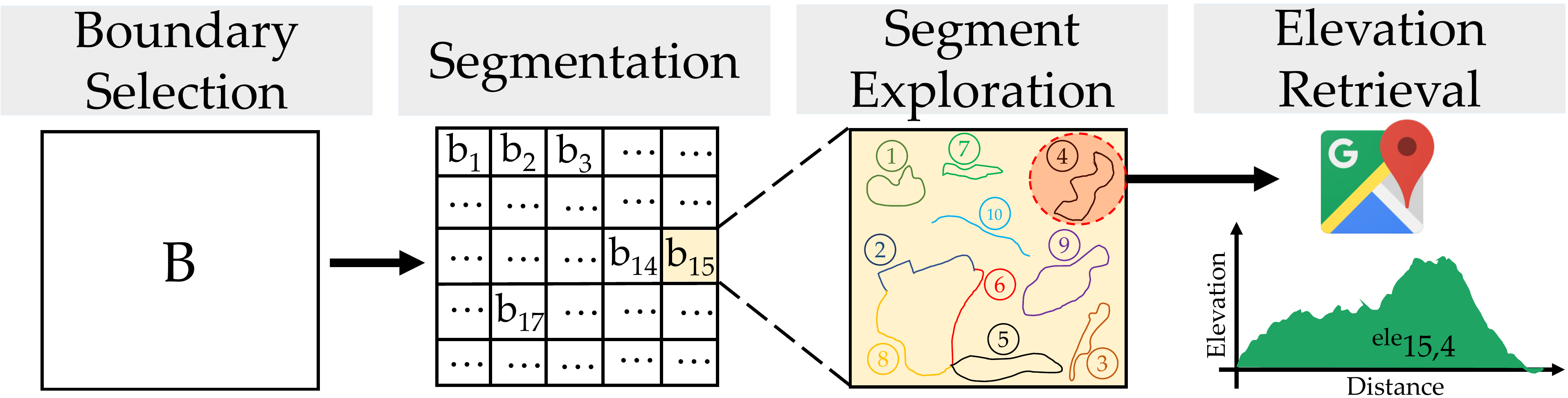}
    \caption{An illustration of the data mining pipeline. A geolocation boundary, B, is segmented into small boundaries, each of which is then forwarded to the Segment Exploration step to obtain the most frequented ten route segments for that particular boundary. Finally, the elevation profile of each route segment is retrieved from the Google Maps API.}
    \label{figure:miningpipeline}
\end{figure}

\subsubsection{City-Level Dataset} 
For the city-level dataset, we mined {\bf publicly available} training route segments in a popular fitness tracking application using its \exploreSegments functionality.
We note that our experiments do not put any users at risk and are not in violation of the terms of use of the fitness tracking application: since both the trajectory (map) and elevation profiles are public information, we are also not obtaining any information beyond what is provided by the users explicitly. 
We also note that the training route segments are user-created activity routes whose main purpose is to compare completion times among users who also completed the same route. 
They are particularly useful for our purposes since they include public location trajectories that are frequented by the actual users rather than randomly created location trajectories that may not necessarily be of privacy value. 
During the segments mining, the anonymity---thus the privacy---of the users who frequented the segments or created the segments is maintained. 

The overall data mining procedure consists of three steps, as illustrated in \autoref{figure:miningpipeline}.
First, we define the cities of interest, which we also use as the class labels per our threat model. 
For each city, we define the rectangle geolocation boundary box $B$ consisting of the top right and bottom left corner coordinates in the boundary selection phase. 
In the segmentation phase, and since \exploreSegments returns only the ten most frequented segments encapsulated by a given boundary, and to obtain more data of a geolocation boundary box, we divide the large rectangle boundary of the city into smaller region boundaries, each denoted by $b_i$, by following a grid-like structure as shown in the second phase of the \autoref{figure:miningpipeline}.
For each region boundary $b_i$, we call \exploreSegments and receive the geolocation polyline path, $path_i^j$ where $ j \in [1,10]$, of the 10 most frequented segments encapsulated in $b_i$, as shown in the segment exploration phase. 
Finally, since the polyline paths do not include elevation profiles, we obtain the associated elevation profile $ ele_{i,j} $ for each $path_i^j$ using the Google Maps Elevation API through the elevation retrieval phase. 
The sample size distribution of the city-level dataset can be found in \autoref{table:city-level}. 

Unlike the user-specific dataset, the city-level dataset does not include any overlapped samples since each region $r_i$ is disjoint with the other regions. 
A segment route included by more than one neighboring region is not considered since \exploreSegments returns the routes encapsulated within the given boundaries, $b_i$. 
    
\begin{table}[t]
    \caption{User-specific dataset sample size distribution.}
    \centering
    \begin{tabular}{|l|c|r|}
         \hline
         \textbf{Regions} & \textbf{Abbreviation} & \textbf{Sample Size} \\
         \hline
         Washington DC & WDC & 366 \\
         \hline
         Orlando & ORL & 232 \\
         \hline
         New York City & NYC & 120 \\
         \hline
         San Diego & SD & 18 \\
         \hline
    \end{tabular}
    \label{table:raw}
\end{table}


\begin{table}[t]
    \caption{City-level dataset sample size distribution.}
    \centering
    \begin{tabular}{|l|c|r|}
        \hline
        \textbf{Regions} & \textbf{Abbreviation}  & \textbf{Sample Size} \\
        \hline
        New York City & NYC & 2437 \\\hline
        Washington DC & WDC & 2129 \\\hline
        San Francisco & SF & 743 \\\hline
        Colorado Springs & CS & 369 \\\hline
        Minneapolis & MIN & 363 \\\hline
        Los Angeles & LA & 280 \\\hline
        New Jersey & NJ & 266 \\\hline
        Duluth & DUL & 156 \\\hline
        Miami & MIA & 94 \\\hline
        Tampa & TAM & 83 \\\hline
    \end{tabular}
    \label{table:city-level}
\end{table}
    
\subsubsection{Borough-Level Dataset}
For the borough-level dataset, we apply a similar mining procedure as we have done with the city-level dataset, using the borough boundaries instead of the city boundaries. 
\autoref{table:borough-level} shows the sample size distribution of the borough-level dataset for different cities.

\begin{table}[t]
    \caption{Borough-level dataset sample size distribution.}
    \centering
    \begin{tabular}{|c|l|r|}
        \hline
        \textbf{City/State} & \textbf{Region} & \textbf{Sample Size} \\
        \hline
        \multirow{4}{*}{ \makecell{\bf Los Angeles \\(LA)} }
        & Downtown & 280 \\ \cline{2-3}
        & Santa Monica & 128 \\\cline{2-3}
         & Chinatown & 46 \\\cline{2-3}
        & Beverly Hills & 38 \\
        \hline
        \multirow{3}{*}{\makecell{\bf Miami \\(MIA)}}
        & Downtown & 67 \\\cline{2-3}
        & Miami Beach & 44\\\cline{2-3}
        & Virginia Key & 18 \\
        \hline
        \multirow{3}{*}{\makecell{ \bf New Jersey\\ (NJ) }}
        & Jersey City & 266 \\\cline{2-3}
        & West New York & 23 \\\cline{2-3}
        & Newark & 28 \\
        \hline
        \multirow{6}{*}{\makecell{\bf New York City\\ (NYC)}}
        & Manhattan & 2437 \\\cline{2-3}
        & Queens & 353 \\\cline{2-3}
        & Brooklyn (South) & 239 \\\cline{2-3}
        & Brooklyn (North) & 205 \\\cline{2-3}
        & Bronx & 142 \\\cline{2-3}
        & Staten Island & 119 \\
        \hline
        \multirow{4}{*}{\makecell{\bf San Francisco\\ (SF)} }
        & South West & 743 \\\cline{2-3}
        & South East & 144 \\\cline{2-3}
        & North West & 130 \\\cline{2-3}
        & North East & 86 \\
        \hline
        \multirow{2}{*}{\makecell{\bf Washington DC\\ (WDC) }}
        & District of Columbia & 2129 \\\cline{2-3}
        & Baltimore & 218 \\
        \hline
    \end{tabular}
    \label{table:borough-level}
\end{table}

\subsection{Preprocessing}
A key design element in our pipeline is the representation modality of the elevation profile, which will significantly impact the performance of our elevation-location mapping, as we show later. 
We transform the samples into text-like and image-like representations to facilitate feature extraction and feed them into our classification module. 
In this section, we describe the details of the utilized preprocessing methods.

\subsubsection{Text-like Representation}     
\begin{figure}[t]
\centering
\includegraphics[width=0.5\textwidth]{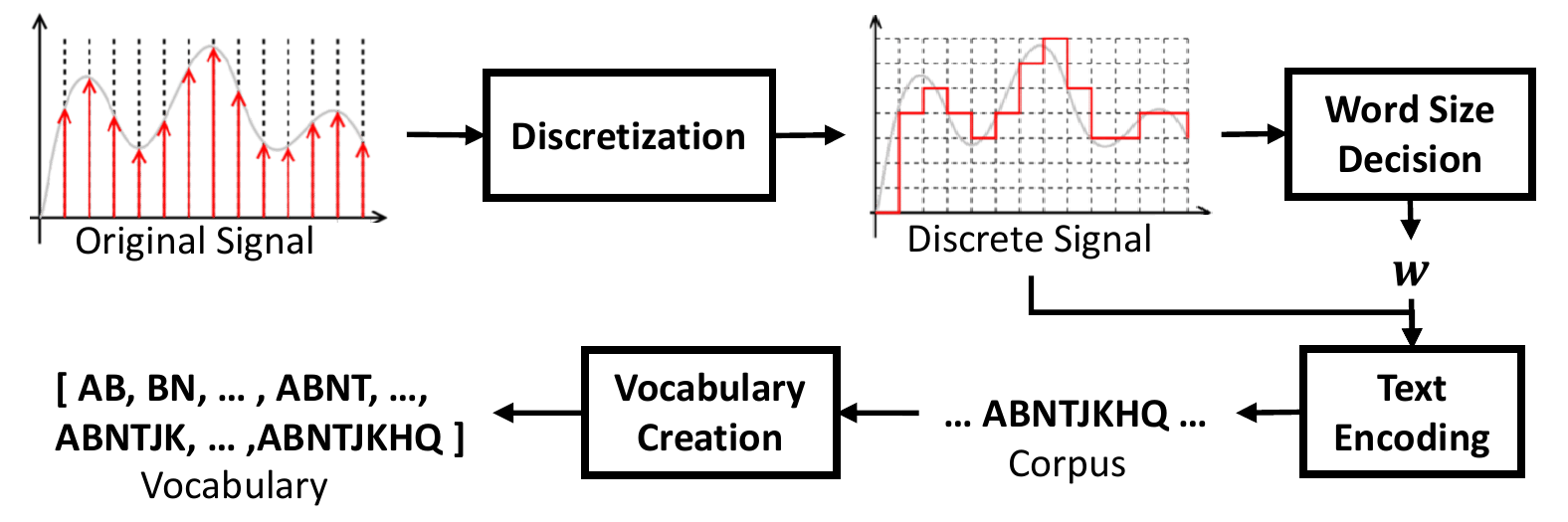}
\caption{Illustration of the flow of text-like preprocessing. The signal is discretized by eliminating the small elevation fluctuations. The discretized signal is also used for deciding the word size of the encoding. The discrete signal is then encoded in text and a vocabulary is built.}
\label{figure:text-preprocessing}
\end{figure}

For our text-like representation, our approach follows four steps, as  shown in \autoref{figure:text-preprocessing}: discretization, word size decision, text encoding, and vocabulary creation. 

\begin{figure}[t]
\centering
\includegraphics[width=0.5\textwidth]{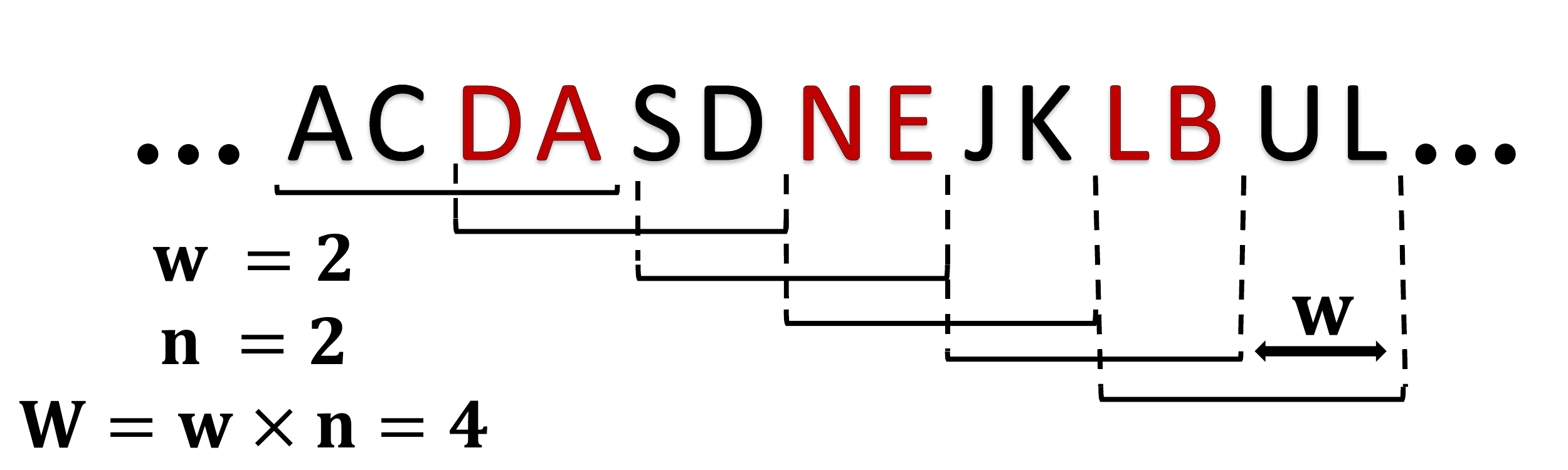}
\caption{Illustration of bi-gram creation where the word size is $w = 2$ and window size is $W = 4$. }
\label{figure:n-gram}
\end{figure}

\begin{description}
\item[Discretization.] In the discretization step, the original elevation signal is discretized along the \textit{y}-axis, which represents the elevation values to avoid possible overhead by small differences in the precision causing longer string correspondences and, consequently, longer vocabulary and sparse feature vectors.
The discretization is done as follows. 
Let $e_i^j$ be the $i$-th elevation value in $j$-{th} sample. The discretization functions are defined as $f( e_i^j ) = \lfloor{} e_i^j \rfloor{}$ and $f( e_i^j ) = \frac{\lfloor{} e_i^j \times 10^3 \rfloor{}}{10^3}$, where the first function is used for processing the user-specific dataset and the second function is used for processing the city-level and borough-level datasets. 
Since the user-specific dataset is dense in terms of sampling rate, using the floor function is enough to represent the routes. 
However, as the city-level and borough-level datasets are already sparse, losing information is undesired, so we used the second function to represent the elevations that differ in up to 3 decimal digits precision. Having more fine-level details when the samples are sparse would help our models to learn more and thus increase the accuracy. To demonstrate the effect of discretization, we measured the vocabulary size of the smallest class of the User-specific dataset, i.e., San Diego. 
For a class as small as the San Diego class, using the second function results in a vocabulary of size 12,870, and using the first function results in a vocabulary of size 3,155.
Such difference in the vocabulary size demonstrates the effect and necessity of discretization.
\item[Word size decision.] For the word size decision, we use $w = \log_l c$, where $w$ is the word size, $l$ is the length of the alphabet, and $c$ is the number of unique value occurrences in the given signals. 
\item[Text encoding.] For text encoding, each unique value in all the discrete signals is mapped to a unique string with length $w$, and each sample signal is encoded by referring to the string correspondences of each value in the discrete signal and concatenating these strings to construct a long text, \ie corpus.
\item[Vocabulary creation.] To create our \emph{vocabulary}, we consider the corpus created from all encoded signals regardless of labels. 
Each line in the corpus represents a sample signal, and each word in a line represents the text correspondence of an elevation value in the sample signal. 
We build a vocabulary from the unique word-based $n$-grams of the document. 
As illustrated in \autoref{figure:n-gram}, a window with size $ W = w \times n $ is slid throughout the corpus and each window content is appended to the vocabulary set. 
Since the vocabulary set does not contain duplicate entries by definition, we construct the vocabulary consisting of unique $n$-grams of the given dataset after traversing the corpus by $n$ times with different window sizes. 
\end{description}

\if1
\begin{figure}[t]
\centering
\includegraphics[width=0.45\textwidth]{figs/Figure5.pdf}
\caption{Illustration of the flow of feature extraction and selection process. The preprocessed signal, \ie corpus, is converted to a vocabulary consisting of n-grams of the elements in corpus. The feature vector, the frequency of the vocabulary in the corpus (bag-of-words vector), is extracted by using vocabulary and the corpus.}
\label{figure:overall}
\end{figure}
\fi
    
\subsubsection{Image-like Representation}
\begin{figure}[t]
\centering
\begin{subfigure}[New York City B/W]{\includegraphics[width=0.22\textwidth]{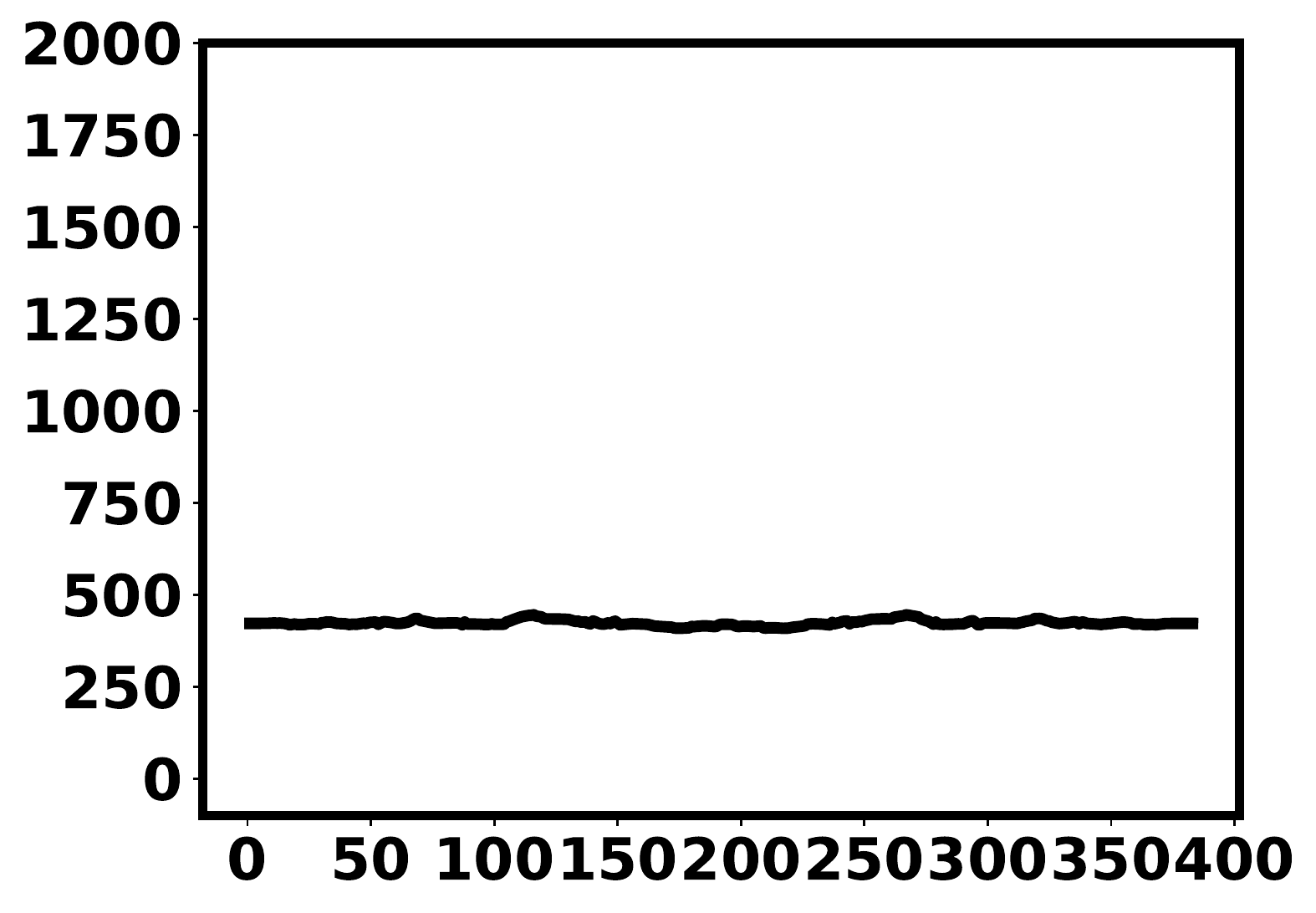}}
\end{subfigure}~
\begin{subfigure}[Los Angeles B/W]{\includegraphics[width=0.22\textwidth]{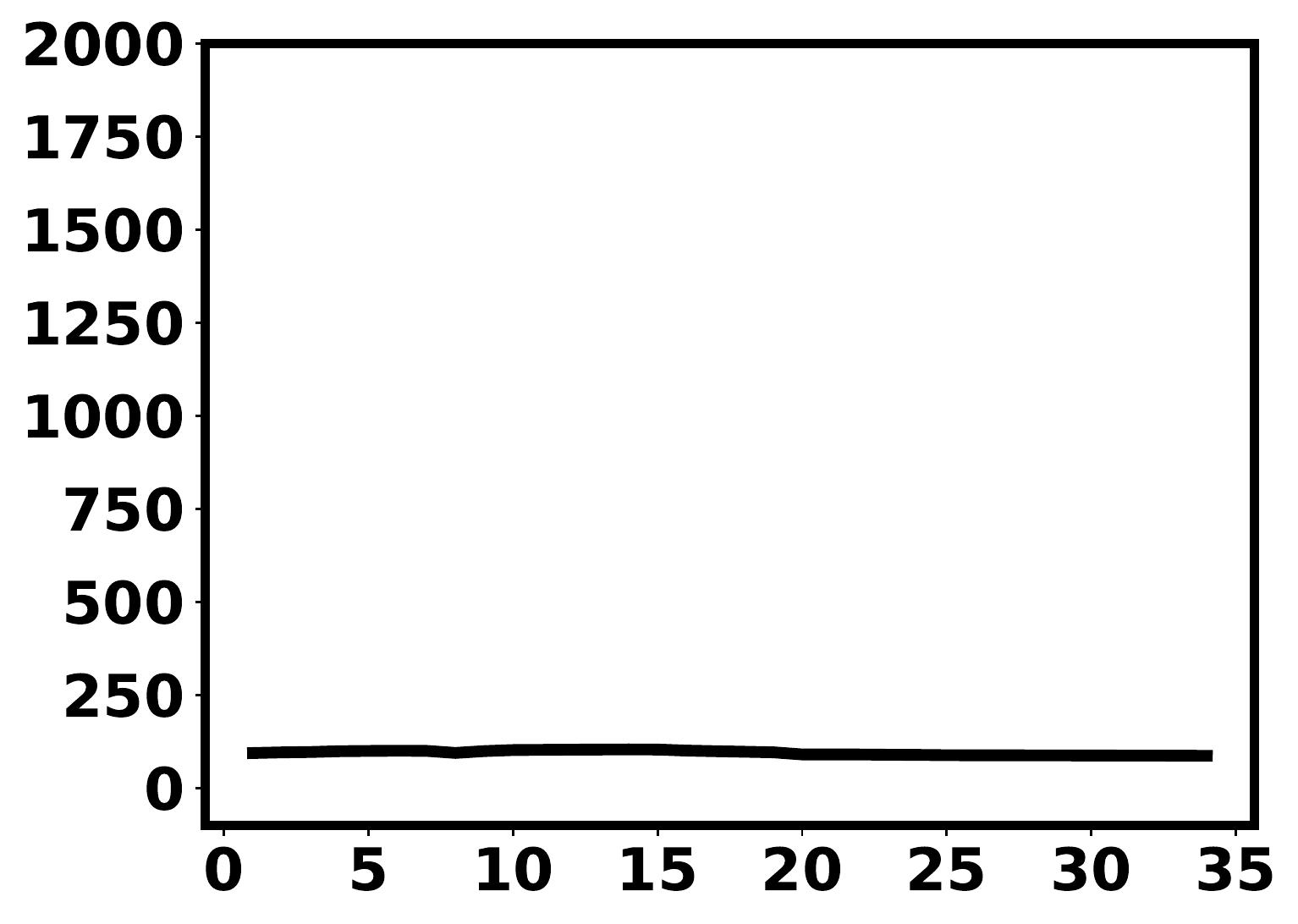}}
\end{subfigure}
    
\begin{subfigure}[New York City colored]{\includegraphics[width=0.22\textwidth]{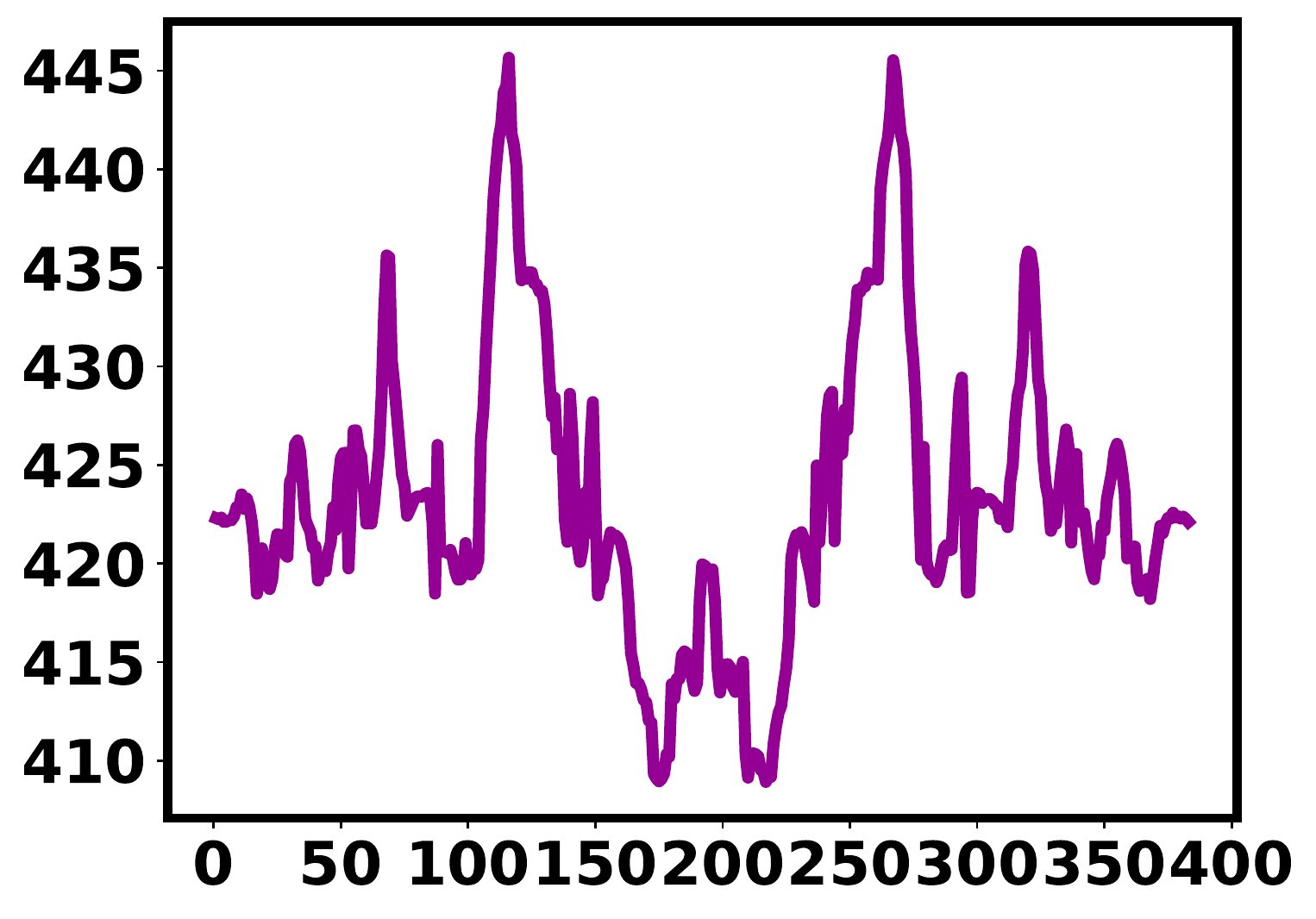}}
\end{subfigure}~
\begin{subfigure}[Los Angeles colored]{\includegraphics[width=0.22\textwidth]{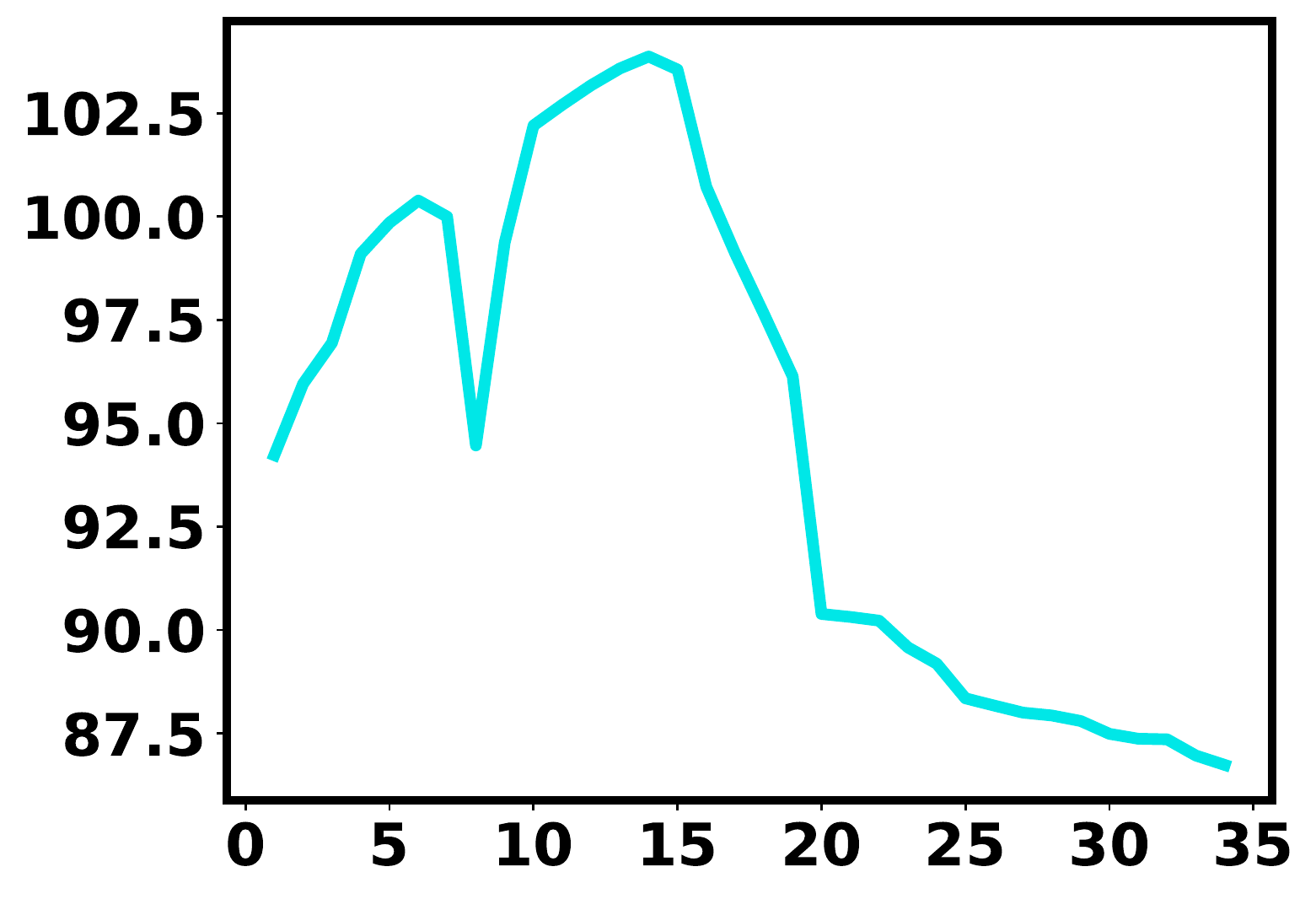}}
\end{subfigure}
\caption{{\textit{Elevation profile graphs by fixing the \textit{y}-axis range and using only black versus elevation profile graphs by fitting and using color encoding.} As can be seen, with the black option the elevation range can be represented by the position of the signal, but changes in the signal are not visible, which is an information loss. However, with the color option enabled, both alterations in the signal and the signal range are represented in one image.} } 
\label{figure:colorless_vs_color}
\end{figure}

In the image-like representation, the elevation signals are drawn as line graphs and saved as a $32 \times 32$ images\footnote{The size is chosen to strike a balance between the performance in terms of the required computations and produced accuracy.}. 
To draw a line graph, the maximum and minimum values for the \textit{y}-axis are set to be the maximum and minimum  of each elevation signal, and the lines are colored to encode the value interval in which the elevation signal ranges. 
We note that the image-like representation with colors has multiple advantages over the black-white representation where the \textit{y}-axis is set to the range of a whole dataset. 
First, as illustrated in \autoref{figure:colorless_vs_color}, the alterations of an elevation signal are more visible with the color encoding method, which could be a more discriminative feature to learn. Second, the color encoding method results in efficient utilization of the feature space.
We use 200 elevation values for each image, obtained by dividing the elevation signal into equal-sized parts.
    
\subsection{Feature Extraction}
To classify elevation profiles accurately, we extract discriminative features from the elevation profile representations. 
    
\BfPara{Text-like} In the text-like feature extraction, we utilized two methods: (i) n-grams, (ii) tf-idf.

For the n-grams method, words and non-overlapping occurrences of word sequences are counted, and a feature vector for each sample is created with each unique word sequence count being a feature. 
Finally, the feature vectors are normalized where each feature represents the occurrence probability of each word in the given sample.

The tf-idf is a statistical feature signifying the importance of a word in a document. 
The tf-idf values proportionally increase as the number of appearances of a word in a document increases. 
Technically, the tf-idf for a word is the multiplication of two metrics: (i) term frequency (tf) and (ii) inverse document frequency (idf). 
Tf-idf of a word $t$ in a document $d$ included in the set of documents $D$ of which cardinality is $N$ is calculated as follows:
\begin{align}
    \nonumber \text{tf-idf}(t,d,D) &= \text{tf}(t,d)~.~\text{idf}(t, D),\\
    \nonumber \text{tf} (t,d) &= \log(1+\text{\sf freq}(t,d)),\\
    \nonumber \text{idf}(t, D) &= \log\left(\frac{N}{\text{\sf count}(d \in D: t\in d)}\right).
\end{align}


The higher the tf-idf means that the word $t$ is more relevant to the particular document $d$.

\BfPara{\em Feature Selection} When the dataset is large and diverse, the vocabulary and, consequently, the feature vector representation become too large to process, compute, and learn from. 
With a feature selection phase to address long feature vectors, some rarely occurring features in the vocabulary are discarded according to a pre-specified feature frequency threshold. For selection, the features are ordered by their term frequency across the corpus, the features whose term frequency is below a specified threshold are discarded, and a new vocabulary is created. 
For both feature extraction methods of the text-like representations, the term frequency threshold is set such that the size of the eventual vector representation is 5,000.

\BfPara{Image-like} \label{Image-like}
We use the CNN for processing and classifying the $32 \times 32$ images, thus it is unnecessary to explicitly extract features since the convolutional layer kernels do that already by learning the filters optimally and efficiently. 
Therefore, the actual feature extraction mechanism for the image-like representation is discussed in the context of the classification phase.

\subsection{Multi-Class Classification}
For classification, SVM, RF, MLP, LSTM, and CNN are used.  

\BfPara{SVM} SVM is a supervised classification technique. 
The main challenge in SVM is finding the best hyperplane that divides the classes from each other considering a given margin. 
SVM with the linear kernel is able to distinguish the classes more successfully when the features are multi-dimensional and numerous~\cite{DBLP:series/faia/2007-160,Hsu2008APG}.
In linear kernel settings, when an optimal hyperplane is found while representing a class, only the features around the hyperplane within the given margin are considered and the other features are simply ignored. 
As such, the complexity of SVM is independent of the number of features. 
Since we consider n-grams up to $n$=5 in the feature extraction phase, the number of features is numerous which makes the usage of SVM legitimate in terms of efficiency and success.

SVM with a linear kernel is generally used for binary classification.
To utilize it for the multi-class classification problem, we use the one-versus-rest method.
In this method, an individual model is trained for each class and the label of the most confident model is outputted.
For the penalization, we use the L2 norm, as it is a standard for linear SVMs.
As for the loss function, we utilized the square of hinge loss, and the hyperparameters are decided through grid search tuning.

\BfPara{RF} RF is based on decision trees, which are ensemble learning methods for classification. 
The main feature of ensemble methods is further improving the generality and robustness of a single estimator by combining several base estimators that are built with a given learning algorithm. 
In decision trees, features are represented by tree nodes and each branch between two nodes represents what the immediate ancestor node returned.
Since building an optimal binary decision tree from given features is an NP-complete problem, using a Random Decision Forest with different tree configurations and efficient heuristics is a way to alleviate the NP-completeness for classification problems. 
While creating our random forest, perturb-and-combine techniques are used. 
Perturb-and-combine techniques are designed specifically for decision trees to improve their accuracy by creating several (different) versions of the estimator by \emph{perturbing} the training set, then \emph{combining} these different versions into a single estimator~\cite{breiman1998ac}. 
Further details on this approach can be found in \cite{Breiman:2001:RF:570181.570182} and \cite{breiman1998ac}. 

In this study, we use 20 decision trees for the RF model.
The final prediction is then done by averaging the tree predictions.
We do not set any upper limit on the number of leaf nodes or the depth of the tree, i.e., there were no time optimization concerns during training the process, so we leave the trees to grow to their maximum depth.

\begin{figure}[t]
\centering
\includegraphics[width=0.5\textwidth]{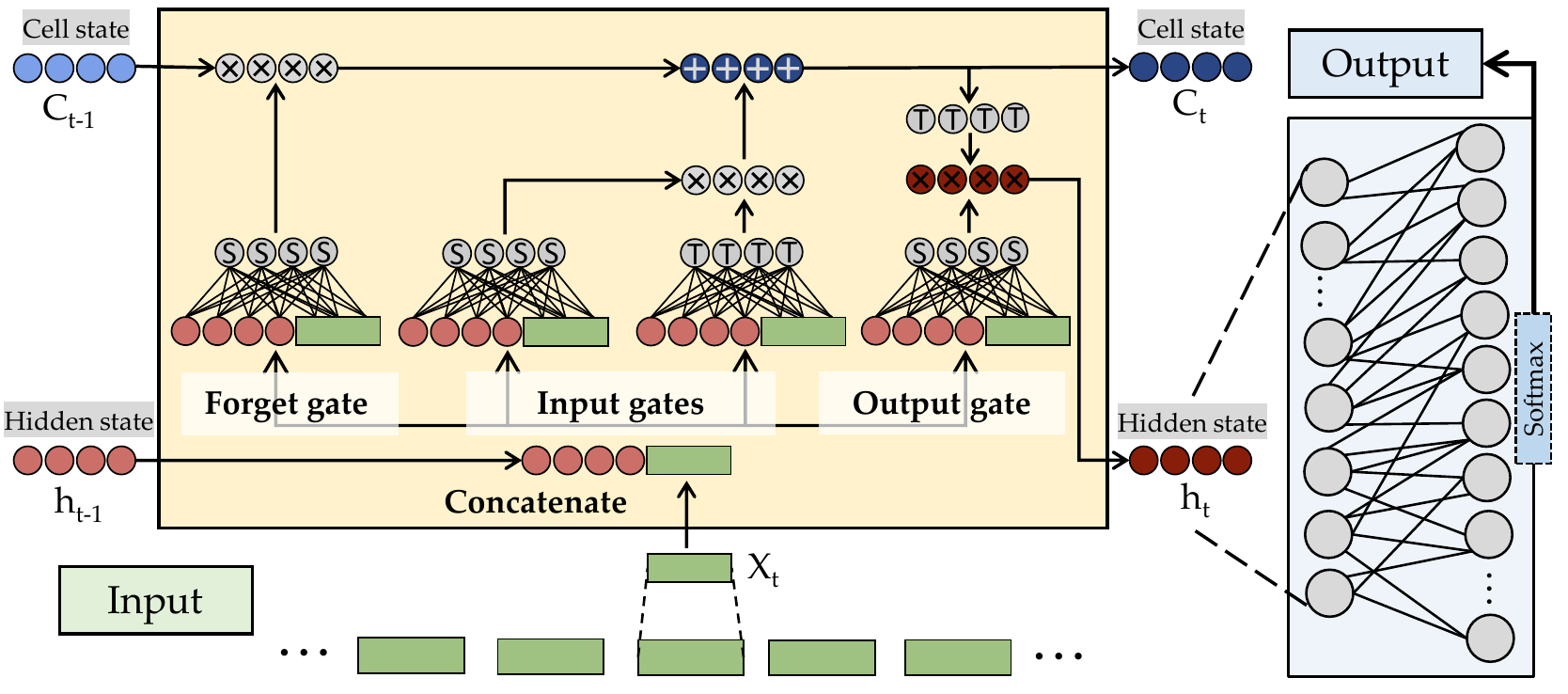}
\caption{The architecture of the LSTM network consists of an LSTM unit with four hidden layers and two fully-connected layers. The input samples are passed through the LSTM unit individually and the last hidden state of the LSTM unit is forwarded to the fully-connected neural network. The fully-connected neural network utilizes softmax activation at the output layer which outputs the class probabilities.}
\label{fig:LSTM_architecture}
\end{figure}

\begin{figure}[t]
\centering
\begin{subfigure}[1-D CNN Architecture \label{figure:CONV1D}]{\includegraphics[width=0.48\textwidth]{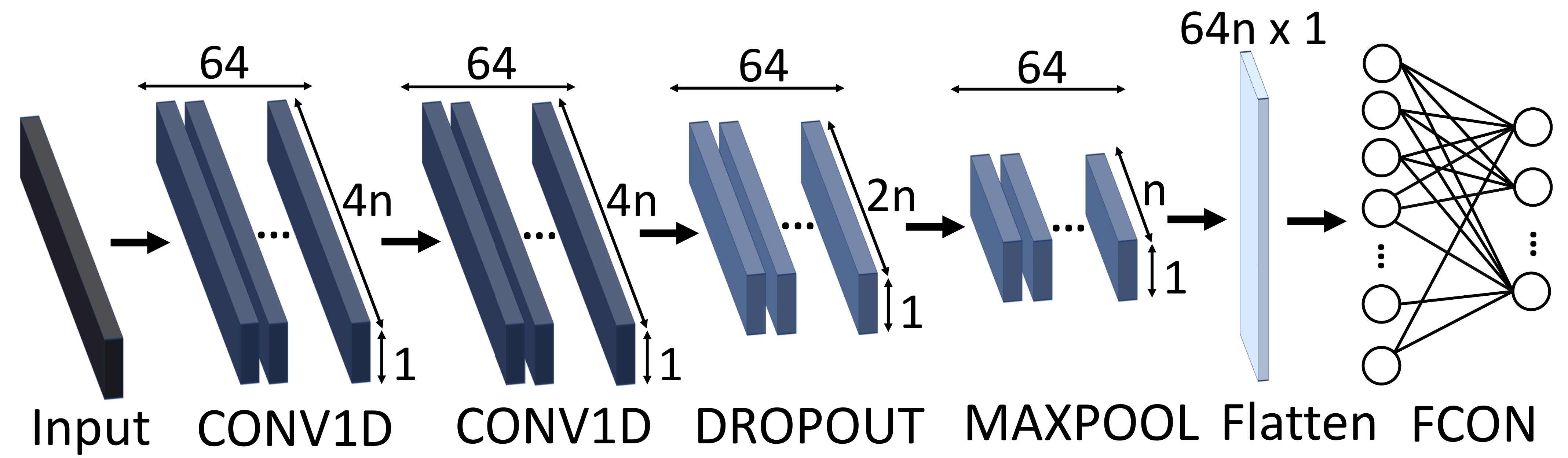}}
\end{subfigure}
\\
\begin{subfigure}[2-D CNN Architecture \label{figure:CONV2D}]{\includegraphics[width=0.5\textwidth]{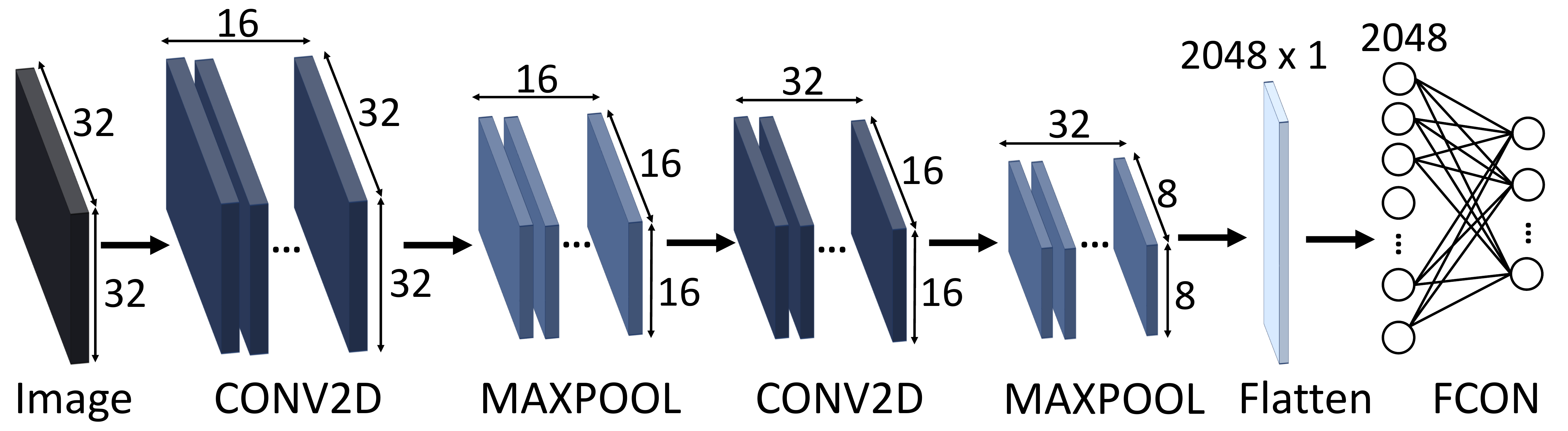}}
\end{subfigure}
\caption{The CNN architectures used for classification. A 1-D CNN is used for the one-dimensional data representations, i.e., n-grams, tf-idf, and raw data. A 2-D CNN is used for the image-like representation. The numbers in the figures depict the dimensions of the input throughout the learning/prediction process.}
\label{figure:CNNS}
\end{figure}

\BfPara{MLP} MLP is a feed-forward fully-connected neural network that utilizes backpropagation for training and is used for supervised learning. 

Recent studies on comparing  multi-layer neural networks and decision trees \cite{Lim2000, Eklund02aperformance} concluded the following:
\begin{itemize}
    \item Multi-layer neural networks allow incremental learning, in which the model's knowledge is continuously extended, more easily than the decision trees.
    \item The training time of the multi-layer neural networks is much longer than decision trees.
    \item The multi-layer neural network predictions are generally as good as the predictions produced by the decision trees, although they can perform better in certain cases.
\end{itemize}

Given the aforementioned potential for MLP to perform better than RF, we used MLP with 20 hidden layers in our experiments. 
We used Adam solver \cite{DBLP:journals/corr/KingmaB14} for weight optimization since it is shown to perform well in terms of both training time and validation score for large feature spaces.
We also utilize ReLU as the activation function, 0.001 as the learning rate, and 200 as the epoch size.
For regularization, we use L2 norm with 0.0001 penalty parameter.
The hyperparameters are decided through grid search tuning.

\BfPara{LSTM} LSTM is a recurrent neural network architecture. 
Unlike the standard neural networks, LSTMs are capable of keeping track of long-term dependencies in the input sequences using feedback connections. 
Such long dependencies are handled through feature extraction in n-grams and tf-idf vectors for standard neural networks. LSTM, on the other hand, handles dependencies internally, without requiring an explicit feature extraction.
LSTM is also particularly useful for capturing the order dependence in sequence prediction problems.
For LSTM, the input sequence can be a time series, a sentence from a given language, or a text-like representation as in our application case.
\autoref{fig:LSTM_architecture} depicts the LSTM architecture employed in this work, where
the input is directly passed through the LSTM layer with four hidden layers, and two fully-connected layers following the LSTM layer.
Each input sample (vector representation of an elevation profile for n-grams and tf-idf, or individual values for the raw data) is passed through the LSTM unit.
When all sample vectors (or elevation values) are passed through the LSTM units, the final hidden state vector is passed to the subsequent fully-connected layer.
In the LSTM unit, hyperbolic tangent and sigmoid (depicted as T and S, respectively, in  \autoref{fig:LSTM_architecture}) are used as the activation function.
For the fully-connected layers, ReLU and softmax activation functions are used, respectively.

\begin{table*}[t]
\centering
\caption{The overall evaluation results for \tma. Accuracy (\%) with different data representations and classification models. In this table, the following abbreviations are used: SVM: Support Vector Machine; RF: Random Forest; MLP: Multi-Layer Perceptron; C1D: 1D Convolution; LSTM: Long Short Term Memory. \textbf{C} column indicates the number of classes in the classification problem. The following settings are used: 4-class = [{WDC, ORL, NYC, SD}], 3-class = [{WDC, ORL, NYC}], 2-class = [{WDC, ORL}]. }
\scalebox{0.95}{
\begin{tabular}{|c|ccccc|ccccc|ccccc|}\hline
\multirow{2}{*}{ \makecell{\bf C }}
& \multicolumn{5}{c|}{\textbf{raw data}} & \multicolumn{5}{c|}{\textbf{n-grams}} & \multicolumn{5}{c|}{\textbf{tf-idf}}            \\ \cline{2-16} 
 & SVM   & RF    & MLP   & C1D    & LSTM  & SVM   & RF    & MLP   & C1D    & LSTM   & SVM   & RF    & MLP   & C1D   & LSTM  \\ \hline
\textbf{2} & 95.29 & 98.43 & 96.60 & 97.45  & 96.61 & 97.35 & 98.22 & 99.11 &  99.74 &  49.67  & 98.89 & 97.56 & 98.89 & \textbf{99.94} & 51.11 \\\hline
\textbf{3} & 77.56 & 98.64 & 95.88 & 96.20  & 96.46 & 96.79 & 97.53 & 97.93 &  99.23 &  45.83 & 98.86 & 97.91 & 98.48 & \textbf{99.00} & 42.60  \\\hline
\textbf{4} & 70.39 & 96.51 & 71.87 & 74.17  & 75.78 & 93.33 & 87.99 & 95.66 &  \textbf{99.80} &  38.67 & 92.33 & 90.66 & 92.33 & 99.54 & 33.33  \\ \hline
\end{tabular}}
\label{table:TM1_text}
\end{table*}

\begin{table*}[t]
\centering
\caption{The overall evaluation results for \tmb. Accuracy (\%) with different data representations and classification models. In the table, we use the following abbreviations: LA: Los Angeles; MIA: Miami; NJ: New Jersey; NYC: New York City; SF: San Francisco; WDC: Washington, D.C. }
\scalebox{0.92}{
\begin{tabular}{|c|ccccc|ccccc|ccccc|}\hline
                       & \multicolumn{5}{c|}{\textbf{raw data}}                   & \multicolumn{5}{c|}{\textbf{n-grams}} & \multicolumn{5}{c|}{\textbf{tf-idf}}    \\ \cline{2-16} 
\textbf{Cities}& SVM & RF    & MLP   & C1D   & LSTM  & SVM   & RF    & MLP   & C1D   & LSTM  & SVM   & RF    & MLP   & C1D   & LSTM  \\ \hline
\textbf{LA}  & 67.18 & 77.41 & 63.43 & 32.50 & 36.88 & \textbf{78.02} & 74.33 & 77.27 & 55.29 & 28.25 & 76.27 & 76.02 & 75.33 & 58.49 & 23.00  \\\hline
\textbf{MIA} & 69.57 & 80.85 & 67.17 & 80.55 & 54.10 & 75.55 & 77.55 & 75.77 & 77.33 & 40.33 & 83.77 & 82.44 & 72.00 & \textbf{99.54} & 25.33  \\\hline
\textbf{NJ}  & 65.56 & 82.56 & 78.31 & 83.32 & 74.18 & 74.76 & 66.19 & 82.39 & 71.19 & 39.05 & 77.93 & 82.69 & 65.71 & \textbf{92.42} & 32.14 \\\hline
\textbf{NYC} & 73.63 & \textbf{84.26} & 73.44 & 37.33 & 25.57 & 82.25 & 80.71 & 79.78 & 73.02 & 20.72 & 81.50 & 82.96 & 82.63 & 76.60 & 17.94  \\\hline
\textbf{SF}  & 65.92 & 74.66 & 65.89 & 42.21 & 32.53 & 74.71 & 76.15 & \textbf{80.25} & 54.08 & 25.88 & 76.13 & 75.71 & 74.71 & 58.07 & 27.92  \\\hline
\textbf{WDC} & 53.08 & 77.30 & 60.44 & 64.01 & 56.05 & 76.13 & 70.63 & 67.20 & \textbf{89.61} & 58.74 & 75.44 & 74.34 & 55.50 & 85.24 & 51.62 \\ \hline
\end{tabular}}
\label{table:TM2_text}
\end{table*}

\begin{table*}[t]
\centering
\caption{The overall evaluation results for \tmc. Accuracy (\%) with different data representations and classification models.  }
\scalebox{0.95}{
\begin{tabular}{|c|ccccc|ccccc|ccccc|}\hline
\multirow{2}{*}{ \makecell{\bf C }}
& \multicolumn{5}{c|}{\textbf{raw data}} & \multicolumn{5}{c|}{\textbf{n-grams}} & \multicolumn{5}{c|}{\textbf{tf-idf}}  \\ \cline{2-16} 
 & SVM   & RF             & MLP   & C1D   & LSTM  & SVM   & RF    & MLP   & C1D   & LSTM & SVM   & RF    & MLP   & C1D   & LSTM \\ \hline
\textbf{3}  & 61.53 & \textbf{88.99} & 64.29 & 82.53 & 56.15 & 76.70 & 78.04 & 77.20 & 65.23 & 33.92 & 85.22 & 81.33 & 81.81 & 65.99 & 33.20 \\\hline
\textbf{5}  & 73.14 & \textbf{93.00} & 73.53 & 65.65 & 70.11 & 80.33 & 78.67 & 79.53 & 53.92 & 20.28 & 86.18 & 82.78 & 84.11 & 52.38 & 20.11 \\\hline
\textbf{7}  & 77.58 & \textbf{94.94} & 80.60 & 52.58 & 64.98 & 85.22 & 84.73 & 84.82 & 43.50 & 13.86 & 88.59 & 88.01 & 87.27 & 48.14 & 14.23 \\\hline
\textbf{8}  & 80.60 & \textbf{95.98} & 80.50 & 44.57 & 67.55 & 84.77 & 84.85 & 85.12 & 43.79 & 14.03 & 87.19 & 86.24 & 86.55 & 50.28 & 12.50 \\\hline
\textbf{10} & 83.72 & \textbf{95.36} & 84.11 & 40.81 & 59.20 & 87.46 & 87.78 & 87.12 & 31.22 & 16.95 & 89.48 & 87.99 & 88.41 & 43.86 & 12.56 \\ \hline
\end{tabular}}
\label{table:TM3_text}
\end{table*}

\BfPara{CNN} CNN is similar to neural networks in the mechanism, as both of them consist of neurons with learned parameters, weights, and biases. 
The improvement of CNN, however, is in the form of convolution layers, which apply forward passes that decrease the number of parameters of the neural network considerably.
Convolution layers also facilitate the processing of high-dimensional data, such as images.
They prepare high-dimensional data for a fully-connected layer, which cannot process high-dimensional data efficiently, by highlighting the important spatial features along the way.

In this study, we utilize two CNN architectures, differing in the convolution layers dimensions.
\autoref{figure:CNNS} illustrates the employed CNN architectures. 

In the first architecture, we use two consecutive 2D convolution layers (CONV2D) along with the ReLU activation function and MAX pooling layers (MAXPOOL) before a fully connected layer (FCON). 
For both of the convolution layers, kernel, stride, and padding sizes are determined as 5, 1, and 2, respectively, based on the performance. 
The distinctive features are selected at the max-pooling layers with a kernel and a stride size of 2, which reduce the dimensions from ($32 \times 32$) to ($8 \times 8$) at two passes.

In the second architecture, we used two consecutive 1D convolution layers (CONV1D)  along with the ReLU activation function. 
A dropout (DROP) layer is added to alleviate the overfitting problem. 
Then, a max-pooling layer (MAXPOOL) and a fully connected layer (FCON) are added.

For both architectures, the softmax function is used as an activation function at the output layer and the Categorical Cross Entropy is used as a loss function. 
For parameters optimization, we used the Adam optimizer.

\section{Evaluation Results}\label{sec:results}

We performed experiments for each dataset, data representation, and threat model.
We categorize the evaluations into three: Raw, Text-like, and Image-like.

\BfPara{Raw} First, we performed evaluations on the raw data. As the multi-class classification models require a fixed input shape and an arbitrary elevation profile does not have a specific length, we divided the elevation profiles into equal-length (32) chunks and use the raw data to train and test the models.
For all datasets and threat models, we used a slightly modified version of the soft voting ensemble method while testing with raw data. 
In conventional ensemble learning techniques, the input is passed to different models and the final prediction is assigned based on the decisions of the models. However, in this study, we passed the equal-length chunks of a single input, i.e. elevation profile, to a single model, and then assigned the final prediction based on the decisions on the input chunks.
Soft voting sums up the predicted probabilities for each class label for each input chunk and returns the class label with the highest probability as the final prediction.

\BfPara{Text-like}
Second, we performed evaluations with text-like features: n-gram, and tf-idf.
With n-gram features, we performed experiments using 10-fold cross-validation and by fixing the dimension of $n$-grams to 5 for all datasets and associated threat models. 
With tf-idf features, we performed 10-fold cross-validation and fixed the dimension of $n$-grams to 5 for all datasets and threat models.

The user-specific dataset contains overlapped and repetitive portions by nature. 
In the Simulations subsection, we simulated the same behavior on the mined datasets and performed the same evaluations for comparison. 

\BfPara{Image-like}
For the experiments on the image-like representations, we employed three methods in CNN: unweighted loss function, weighted loss function, and fine-tuning. 
In the unweighted and weighted loss function evaluations, we split the test data from the dataset by considering the sample size of the classes; we assigned probabilities for each class considering the inverse proportion to its size and then randomly selected test data with the associated probabilities. 
In fine-tuning evaluations, we performed 10-fold cross-validation in the last round where all the classes have the same sample size. 

\subsection{Raw and Text-like Data Evaluation}

\subsubsection{Direct Evaluation} 

\BfPara{\cib{1} Evaluating \tma} We trained and tested models with the user-specific dataset.
As shown in \autoref{table:raw}, the user-specific dataset has an unbalanced sample size across classes. 
To mitigate bias, we use the same sample size for each class and change the number of classes at each step.
The evaluation results are shown in \autoref{table:TM1_text}. 
Due to the limited number of samples, the accuracy decreases as the number of classes increases.
The only exception is C1D with n-grams and tf-idf.
One-dimensional convolutions were able to capture the characteristics of elevation profiles even with a limited number of samples.
The results show $ 99.80\% $ accuracy with C1D, n-grams, and 4-class classification. 
With tf-idf and C1D, we obtained $ 99.00\% $ and $ 99.94\% $ accuracy with 3-class and 2-class classification, respectively. 

LSTM gives better accuracy with raw data compared to the other text-like representations. 
LSTM performs better on the data where the ordering is decisive.
With the text-like representations, the original ordering of the values is encoded separately, thus LSTM could not extract much information through the ordering.

For \tma, RF also performs better with raw data.
Since extracting features from the range and the ordering of the values are less demanding for decision trees, it is reasonable to observe such a pattern.

Other classification methods, i.e., SVM, MLP, and C1D, benefit more from the n-grams and tf-idf features.

Since the user-specific dataset is compiled from actual users, exhibiting mobility patterns, about $35\%$ of the routes are overlapped. 
In a repetitive and overlapped setting, both training and testing splits may contain similar patterns leading to high accuracy scores. 
The results prove that a targeted attack on a person whose activity history is known will be successful with accuracy as high as $99.80\%$. 

\BfPara{\cib{2} Evaluating \tmb}
While evaluating \tmb, the borough-level dataset is used. 
Individual models are created for each of the cities, by labeling the data as the name of the corresponding borough and evaluated separately. 
Similar to the user-specific dataset, the borough-level dataset also has an unbalanced sample size across the classes.
To avoid biased results, we fix the sample size to that of the smallest class for all classes. 
At each fold, we randomly select train and test data for the classes with more samples.
\autoref{table:TM2_text} shows the accuracy results of each model. 

\textbf{Los Angeles} model reaches up to $78.02\%$ accuracy with n-grams and SVM.
Similar results are obtained with other classification methods and data representation pairs, such as raw and RF, n-grams and MLP, tf-idf, and SVM.
For Los Angeles, C1D could not become prominent; the less complex models perform better on this dataset.

With \textbf{Miami}, we reach up to $99.54\%$ accuracy with tf-idf and C1D.
Overall, tf-idf is shown to be a better representation for this particular dataset.
Combining a complex model with a representative feature, we achieved high accuracy.

For \textbf{New Jersey}, we achieve $92.43\%$ accuracy with tf-idf and C1D.
According to the results, tf-idf features better represent the New Jersey dataset. 

In \textbf{New York City}, we reach up to $84.26\%$ accuracy with raw data and RF. 
When we examine the dataset, we observed that the elevations fluctuate mostly between 13 ft and 95 ft.
When such a small range is considered, decimal digit precision plays an important role. 
Since we do not discard any precision in the raw dataset, it is reasonable to have better accuracy with raw data.
Although the highest accuracy is obtained with raw data and RF, tf-idf is a better choice for other classification methods.

In \textbf{San Francisco}, we achieve $80.25\%$ accuracy with n-grams and MLP.
Both text-like representations present similar accuracy patterns.

For \textbf{Washington DC}, we obtain $89.61\%$ accuracy with n-grams and C1D.
For both text-like representations, C1D shows better performance than other methods.

Overall, we can clearly observe the difference between \tma results and \tmb results.
The two main reasons for this performance gap are that (i) there are no overlapped or repetitive routes among the mined segments in the borough-level dataset, and (ii) the elevation differences and elevation sequences are not distinctive enough within a city to decide in which borough is the given test data is.
The results of the simulated behavior will be discussed in the simulations subsection.

\BfPara{\cib{3} Evaluating \tmc}
In \tmc evaluations, due to sample size differences across the labels in the city-level dataset, we follow the same procedure in \tma evaluations. 
A fixed number of samples is randomly selected from each class for training and testing. 
\autoref{table:TM3_text} shows the results of the evaluation.
Per the reported results, we are able to predict the city of an elevation profile among 10 cities with an accuracy of $95.36\%$, among 8 cities with an accuracy of $95.98\%$, among 7 cities with an accuracy of $94.94\%$, among 5 cities with an accuracy of $93.00\%$, and among 3 cities with an accuracy of $88.99\%$.
For \tmc, for all number of classes, we find the best performing configuration as raw data and RF.
When we look into the reason for the fact that raw data with RF outperforms every other configuration, we observe that the elevation range of the different classes in this dataset plays an important role, similar to \tmb: NYC.
Decision trees in RF are able to capture the features firsthand, without any representation needed in the middle.

When we consider the text-like representations, we observe that tf-idf features better represent the dataset.
The success of the city-level estimations, when compared to the borough-level estimations (\tmb), is due to the elevation range and sequence differences across cities, which is reasonable, even though the dataset is mined in a similar fashion as in the borough-level dataset. 
This mining indicates that the city-level dataset also does not contain comprehensive, repetitive, and overlapped samples. 
The results of the simulated evaluation will be discussed in the following.

\begin{figure*}[t]
\centering
\begin{subfigure}[Simulation of \tmb with n-gram features  \label{figure:SimulatedTM2n-gram}]{\includegraphics[width=0.31\textwidth]{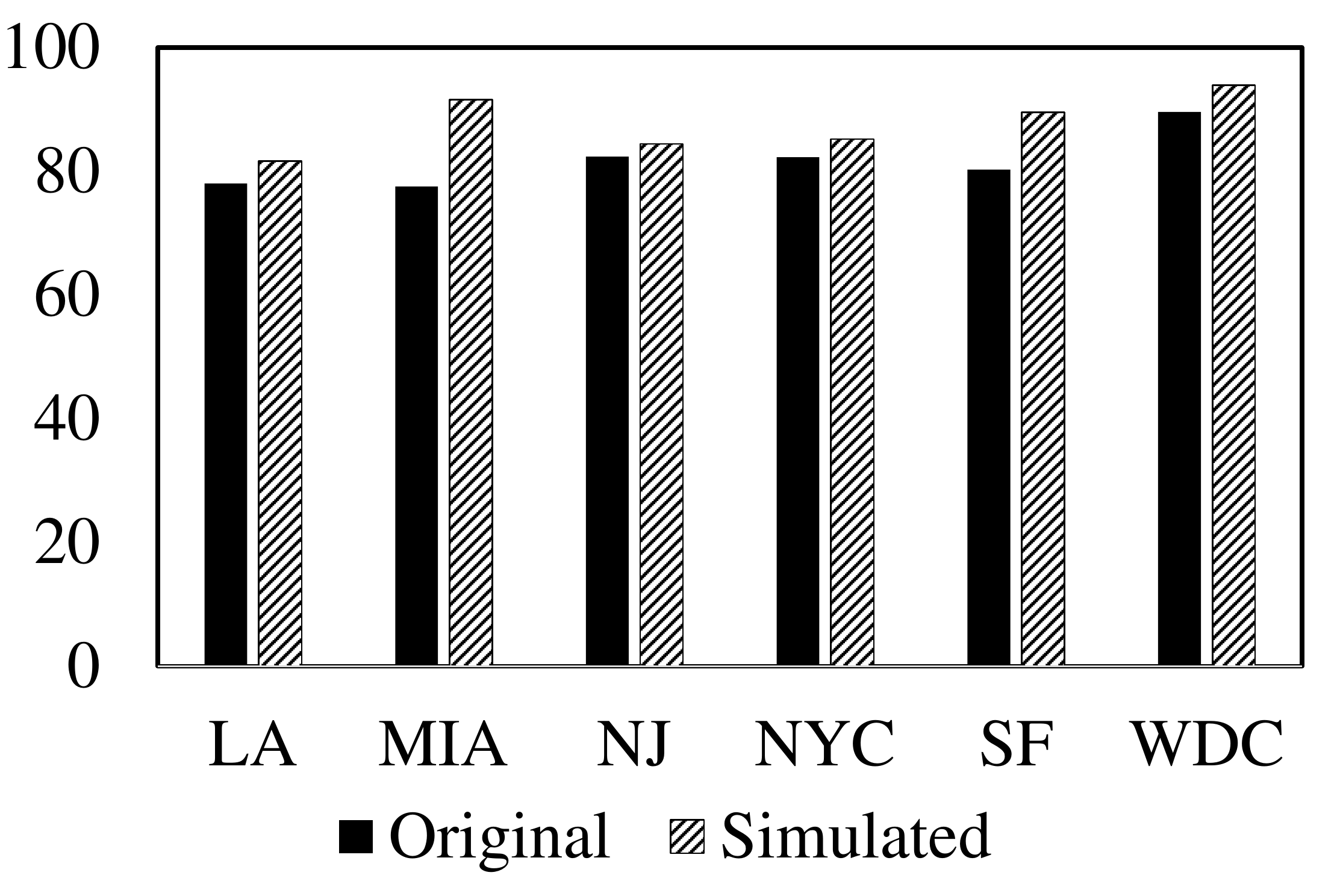}}
\end{subfigure}
\begin{subfigure}[Simulation of \tmb with tf-idf features \label{figure:SimulatedTM2tf-idf}]{\includegraphics[width=0.31\textwidth]{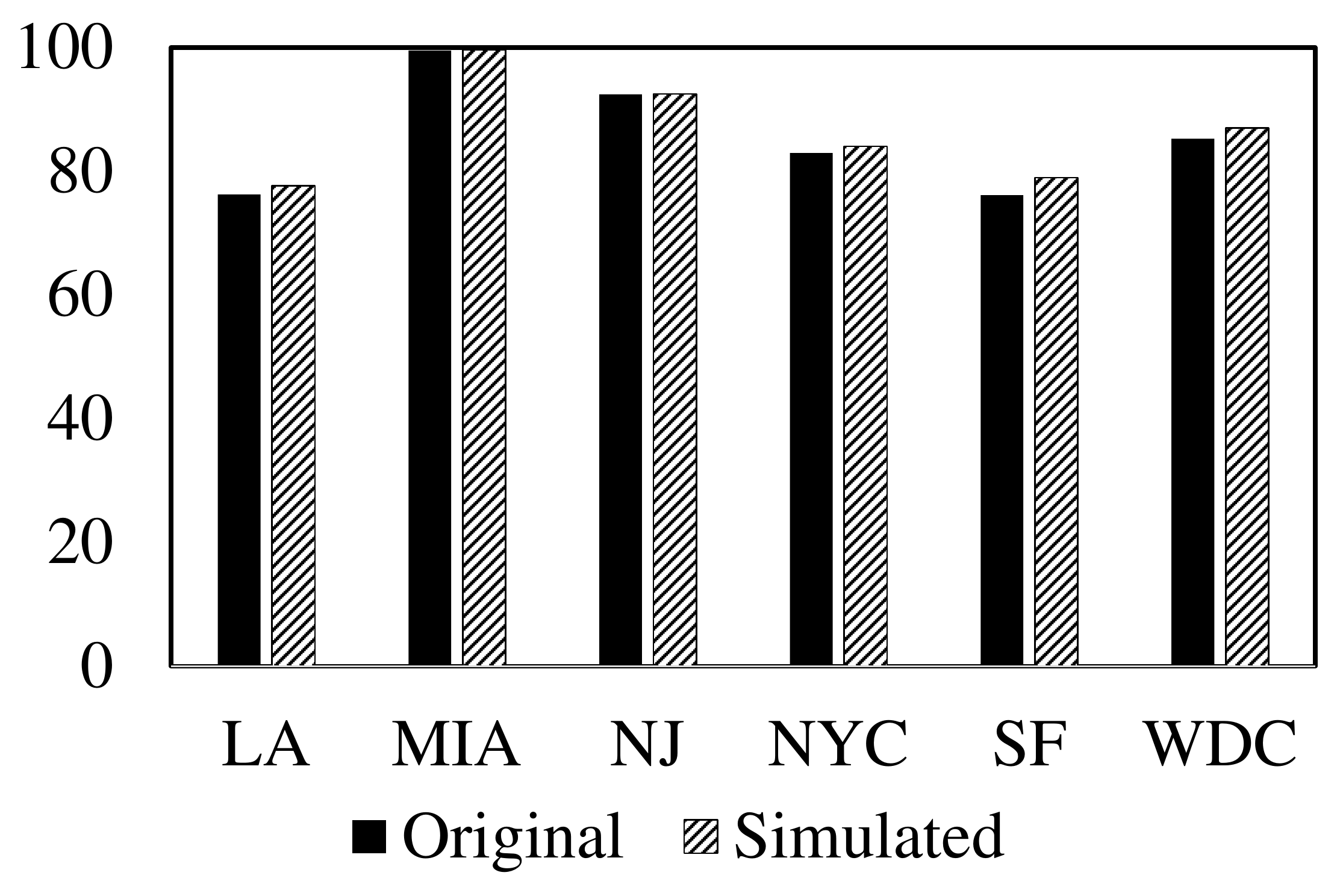}}
\end{subfigure}
\begin{subfigure}[Simulation of \tmc \label{figure:SimulatedTM3}]{\includegraphics[width=0.31\textwidth]{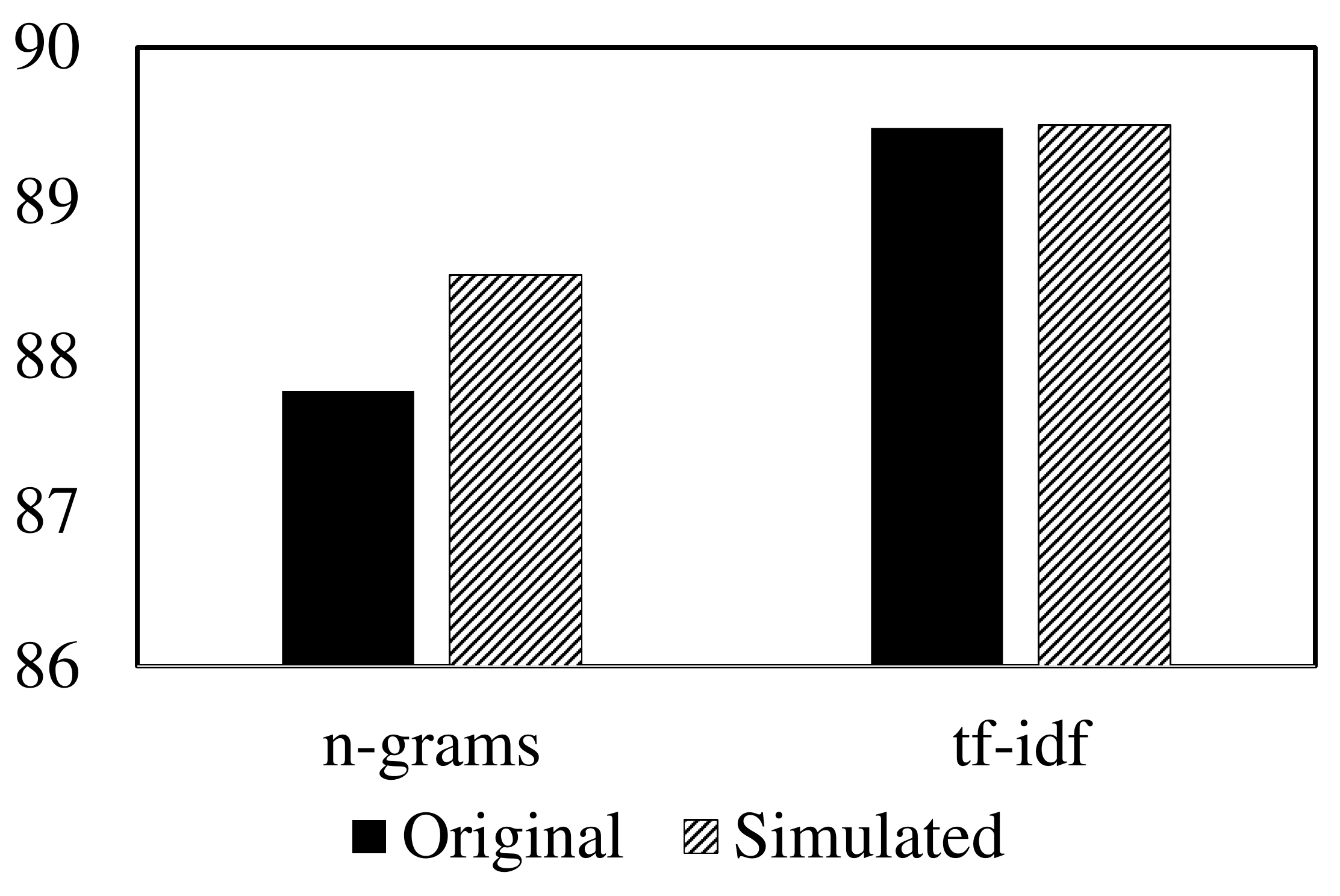}}
\end{subfigure}
\caption{Some selected simulation results of \tmb and \tmc. The maximum achieved accuracy results are compared.} 
\label{figure:simulations}
\end{figure*}

%
\subsubsection{Simulations}
The mined datasets do not contain overlapped or duplicate samples as in the user-specific dataset. 
In this set of evaluations, we simulate overlapped mined datasets and perform evaluations under the same threat models. 

\BfPara{Simulation of \tmb} For the city-level estimation evaluations, we rebuild a simulation dataset with a $30 - 34\%$ overlap ratio for each region within the cities. 
The same evaluation procedures are then followed as the original mined dataset, which is 10-fold cross-validation with a fixed $n$-grams size of 5. 
\autoref{figure:SimulatedTM2n-gram} and \autoref{figure:SimulatedTM2tf-idf} show the comparison between the best-achieved result in the original evaluation and the best-achieved result in the simulations. 
The increase in the accuracy confirms our previous hypothesis that having overlapped route samples would increase the accuracy. 
Since the mined dataset is not specific to any target user's mobility pattern, it is anticipated to result in less accuracy than the \tma evaluation accuracy scores.

\BfPara{Simulation of \tmc} For \tmc's simulated evaluations, we rebuild a simulation dataset with a $35\%$ overlap ratio for each city and performed the same evaluation with 10-fold cross-validation and 5-grams. 
\autoref{figure:SimulatedTM3} shows the comparison of the best-achieved accuracy results in original evaluations and simulations. 
As expected, the accuracy is increased in the simulations proving our previous hypothesis that having similar patterns in a dataset affects the success of the attack.

\subsection{Image-like Data Evaluations}
In this set of evaluations, we perform experiments on the image-like representations of the data. 
Since the original data is unbalanced, the dataset built with the image-like representation also inherits the problem.
In this section, we explain the methods to avoid bias due to an unbalanced dataset and discuss the associated results.

\BfPara{Dealing with Unbalanced Dataset}
There are various methods to deal with unbalanced datasets, including downsampling, oversampling, and creating synthetic samples from existing ones. 
Among these methods, downsampling and oversampling are the easiest ones to explore, although downsampling leads to losing a great amount of data, and oversampling raises the chances of getting lower accuracy as the misclassified duplicated samples increase the false ratio.
Therefore, we explore other alternatives: (i) weighted loss function and (ii) fine-tuning with different samples.

\BfPara{Weighted Loss Function} For the unbalanced dataset, we utilize a weighted loss function while training the CNN and use all the data in the dataset. 
By assigning a class weight that is inversely proportional to the sample size of the class, we signify samples of small classes while calculating the loss, thus their effect does not easily wear off. 

\begin{figure}[t]
    \centering
    \includegraphics[width=0.49\textwidth]{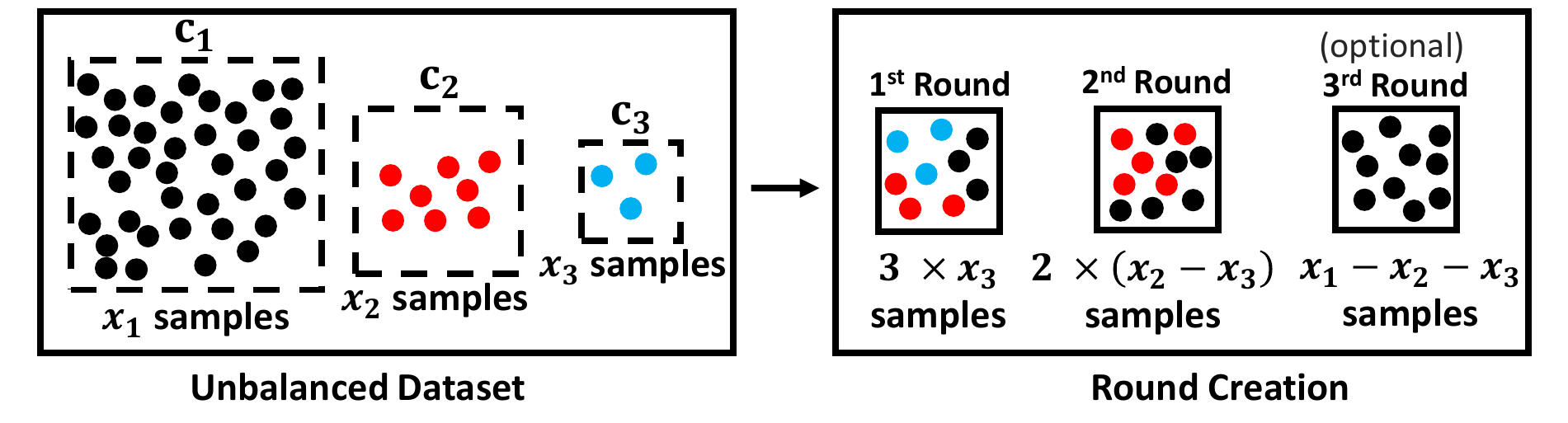}
    \caption{An illustration of round creation from an unbalanced dataset of three classes. }
    \label{figure:round_creation}
\end{figure}

\begin{figure}[t]
    \centering
    \includegraphics[width=0.49\textwidth]{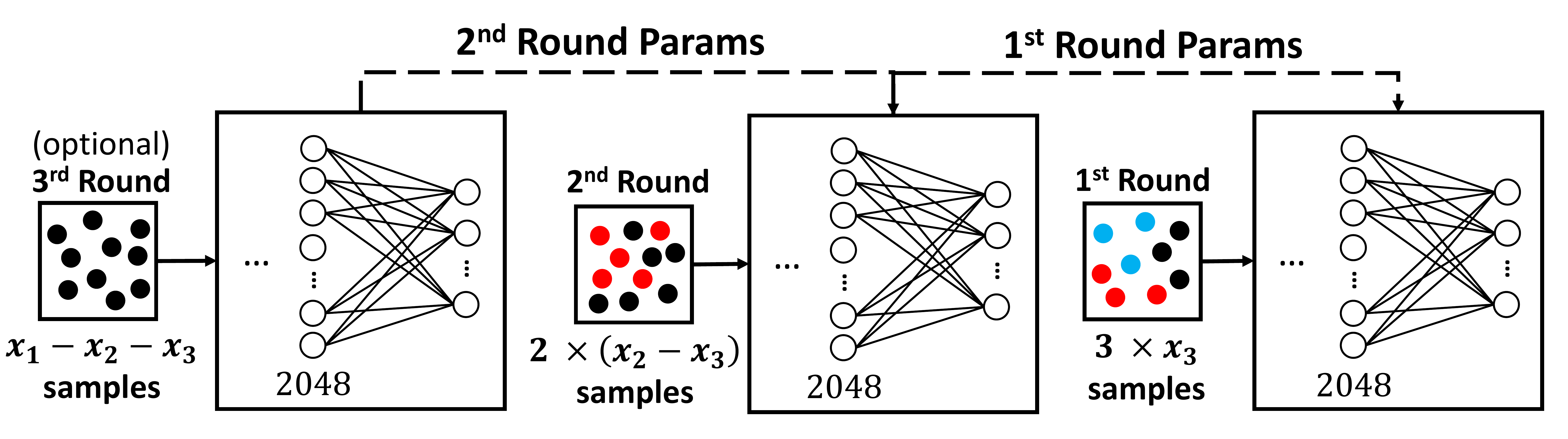}
    \caption{An illustration of the fine-tuning pipeline for an unbalanced dataset of three classes.}
    \label{figure:fine-tuning}
\end{figure}

\BfPara{Fine-Tuning with Different Samples} Fine-tuning is a common technique in deep learning and is used for re-training a complex pretrained model with another dataset. 
To address the unbalanced dataset, we take advantage of fine-tuning in a different manner. 
Namely, we introduce rounds and create a set of small datasets from the unbalanced datasets for each round. 
As illustrated in \autoref{figure:round_creation}, several small and balanced datasets are created by randomly selecting samples. 
For each consecutive round, samples of one or more classes are discarded, and the round dataset is created from the remaining classes. 
After round dataset creation, the model is trained with the round dataset that contains \emph{the least number of classes}, \ie the lattermost created round dataset.
At each step, the model is re-trained using the same or different hyperparameters until all the rounds expire. 
The dataset ordering of the rounds is reversed since the impact of the smallest dataset would wear off if the model is trained with the same order of round dataset creation, which conflicts with the whole idea. 
As illustrated in \autoref{figure:fine-tuning}, while re-training, the parameters of the previous model are passed to the model of the next round. 
The hyperparameters of each round can be tuned accordingly. 
For instance, for the last round, where we include all of the classes, the learning rate is reduced to find the loss minima.

\begin{table}[t]
\centering
\caption{Comparison of maximum achieved accuracy across different methods. The Unweighted Loss (UWL) column is not considered while deciding the maximum accuracy, as the results are biased. The maximum accuracy of each evaluation is written \textbf{bold}, the results that are not considered are written \textit{italic}.}
\scalebox{0.95}{
\begin{tabular}{|l|c|c|c|c|c|}
     \hline
     & \multicolumn{1}{c|}{\textbf{Raw}} & \multicolumn{1}{c|}{\textbf{Text-like}} & \multicolumn{3}{c|}{\textbf{Image-like}} \\
     \cline{2-6}
     Methods &  & \makecell{n-grams/ \\ tf-idf }  & \makecell{UWL \\ \textit{(biased)}} & WL & FT \\
     \hline
     \tma & 96.51 & \textbf{99.94} & \textit{96.98} & 95.23 & 87.93 \\
     \hline
     \tmb: LA   & 77.41 & \textbf{78.02} & \textit{68.85}  & 68.39  & 63.63 \\ \hline
     \tmb: MIA  & 80.85 & \textbf{99.54} & \textit{88.96}  & 86.80  & 62.50 \\ \hline
     \tmb: NJ   & 83.32 & \textbf{92.42} & \textit{93.45}  & 79.42  & 57.14 \\\hline
     \tmb: NYC  & \textbf{84.26} & 82.96 & \textit{74.20}  & 79.37  & 72.79 \\\hline
     \tmb: SF   & 74.66 & \textbf{80.25} & \textit{67.20}  & 78.70  & 65.38 \\\hline
     \tmb: WDC  & 77.30 & \textbf{89.61} & \textit{62.79}  & 70.28  & 71.50 \\
     \hline
     \tmc & \textbf{95.36} & 89.48 & \textit{92.51} & 92.82 & 89.00 \\
     \hline
\end{tabular}}
\label{table:Comparison}
\end{table}


To evaluate our attacks on the image-like data, the elevation profiles are converted into a dataset of images and rounds using the configurations and steps discussed above. 
\autoref{table:Comparison} highlights the maximum achieved prediction accuracy along with comparisons with other methods.

\BfPara{Weighted vs. Unweighted Loss Function} 
To observe the impact of the weighted loss function, we conduct evaluations without giving any weight to the classes in the loss function while using an unbalanced dataset. 
We note that the unweighted loss function evaluation results are biased due to the unbalanced dataset. 
\autoref{table:Comparison} shows the maximum achieved accuracy for each dataset and method. 
Even though the weighted loss function evaluation results are biased, which \emph{seems} successful in outputting the largest class used during training and testing, the biased results remain behind 4 evaluations out of 8. 
In \tma and \tmc, the accuracy scores of unweighted and weighted loss functions are considerably close. 
Thus, we conclude that the weighted loss function improved the prediction performance primarily for \tmb.


\begin{table}[t]
\centering
\caption{ The fine-tuning results for \tma and \tmc as the epoch size changes. }
\scalebox{0.95}{
\begin{tabular}{|c|c|c|c|c|c|c|}
     \hline
     & \multicolumn{3}{c|}{\tma} & \multicolumn{3}{c|}{\tmc} \\
     \cline{2-7}
     \textbf{Epoch Size} & \textbf{500} & \textbf{1000} & \textbf{2000} & \textbf{500} & \textbf{1000} & \textbf{2000} \\
     \hline
     Accuracy & 79.31 & 87.96 & 82.73 & 86.04 & 89.00 & 87.85 \\\hline
     Recall & 55.87 & 67.54 & 63.12 & 29.76 & 45.34 & 38.91  \\\hline
     Specificity & 86.33 & 92.65 & 88.46 & 92.27 & 93.98 & 93.29 \\\hline
     F1 Score & 58.62 & 68.25 & 63.37 & 36.23 & 45.45 & 41.12 \\
     \hline
\end{tabular}
}
\label{table:Fine-tune_results_1_3}
\end{table}

\begin{table}[t]
\centering
\caption{ The fine-tuning results for \tmb as the epoch size is 1000 and learning rate is 0.001 for all rounds. }
\begin{tabular}{|c|c|c|c|c|c|c|}
     \hline
     \textbf{} & \textbf{LA} & \textbf{MIA} & \textbf{NJ} & \textbf{NYC} & \textbf{SF} & \textbf{WDC} \\
     \hline
     Accuracy & 63.61 & 62.52 & 57.13 & 72.84 & 65.34 & 71.55 \\\hline
     Recall & 28.02 & 25.66 & 40.03 & 18.15 & 30.76 & 73.27  \\\hline
     Specificity & 75.84 & 75.97 & 66.75 & 83.43 & 76.35 & 73.22 \\\hline
     F1 Score & 28.83 & 28.64 & 37.55 & 18.46 & 31.47 & 73.44 \\
     \hline
\end{tabular}
\label{table:Fine-tune_results_2}
\end{table}


\BfPara{Fine-tuning vs. Weighted Loss Function} 
For the fine-tuning evaluations, round datasets are created from the original data. 
For \tma, with 4 classes, 3 rounds are created. 
For \tmc, with 10 classes, 5 rounds were created by eliminating 1, 2, 1, and 2 classes at each round, respectively. 
The dataset of \tmb can be considered as a compilation of the dataset of 6 cities: Los Angeles (3 rounds), Miami (3 rounds), New Jersey (2 rounds), New York City (4 rounds), San Francisco (2 rounds), and Washington DC (1 round).
Even though the main idea is to use all the data we have, we decided to downsample the classes with a large sample size. 
For instance, in the evaluation of \tmb: New York City, the biggest class has 5,455 samples whereas the second biggest class has 960 samples. 
In such cases, we did not create an additional round for only one class as this round would have a strong influence over the predictions, \ie overfitting.

\autoref{table:Comparison} shows the fine-tuning method outperformed the weighted loss function method only for \tmb: WDC. 
The difference between the fine-tuning evaluation of Washington DC and others is that we were able to create only one round from the data.
Overall, according to the results shown in \autoref{table:Fine-tune_results_1_3} and \autoref{table:Fine-tune_results_2}, the fine-tuning evaluation is not as successful as the weighted loss function evaluation, since we still lose some data while creating rounds.

\BfPara{Text-like vs. Image-like Evaluations}
When we compare text-like and image-like representations, we can conclude that text-like representation is a better choice for such attacks.
For all evaluations except \tmc, text-like representation outperformed image-like representation.
For \tmc and \tmb:NYC, the raw data and RF configuration is the best choice.

\section{Discussion}\label{sec:discussion}

\BfPara{Defenses} In this study, we are strictly concerned with the elevation profile as a representation of the location. We note that the location itself is the eventual modality of interest, but the exposed information to the adversary from which the adversary will make the inferences is the elevation. To that end, however, the same technique used for location perturbation~\cite{dewri2012local} to defend against location privacy breaches can be applied on the elevation profile, although more straightforwardly and effectively. Namely, two broad classes of defenses could be applied to our problem domain: perturbation and aggregation. 

\BfPara{\em Perturbation as a Defense} To thwart our inference attacks, we can perturb the elevation while maintaining its overall statistical features used for the original application by adding a carefully crafted Gaussian noise driven from the elevation distribution (zero mean and a standard deviation of the original data). We hypothesize that the noise will affect the classifier but will not impact the validity of some of the driven insight from the elevation for the user (e.g., total elevation\footnote{We emphasize that this defense will strictly preserve some but not all of the features of the elevation profile. For instance, by design, the perturbation will not preserve the total ascend and descend, two vital statistical features of the elevation profile.}). To highlight the validity of this approach, we conduct a limited experiment to perturb 10\% of the original raw data used in the original classification problem in \autoref{table:TM1_text}. To keep the elevation signal plausible and convincing, we superimposed dependently generated noise over an epoch of time (10 seconds) and ``clipped'' the generated noise with one standard deviation of a moving average in the current segment\footnote{We note that this plausibility step further restricts the perturbation to make the perturbed elevation profile acceptable by humans. This restriction explains the limited performance of the defense as unrestricted perturbation reduces the accuracy of the inference attack under the same settings (i.e., 10\% perturbation, SVM, C=12) to 43.87\%.}. As a result, we were able to reduce the accuracy from 95.29\% (in the case of SVM, C=2) to 72.46\%. Similar performance degradation is observed with the n-grams and tf-idf in the same settings: by generating those features from the perturbed data, we were able to achieve an accuracy of 67.19\% and 70.31\% with the n-grams and tf-idf, respectively. While not totally subverting the inference, the approach shows an initial promising direction. 

\BfPara{\em Aggregation as a Defense} Another defense is realized by aggregating the elevation profile information into application-compliant statistics, e.g., total ascend, descend, mean ascend and descend, their standard deviation, minimum or maximum (over quantized elevation profile) or sampled elevation signal, which would reduce the effectiveness of our inference attack or block it altogether. The objective of the defense can be realized by rounding (10s of feet of elevation) so that the number of possible segments associated with the given statistical features is large enough to provide plausible deniability through anonymity  (e.g., the correlation between the aggregate and location is weakened). 

\BfPara{\em Feature-level Defenses and Caveats} We note that other approaches that directly perturb the feature representation modality (e.g., images~\cite{oh2017adversarial}) may not be as practical, since the image itself is a constrained domain, and not every pixel in the image domain is a valid perturbation candidate. We emphasize, however, that this issue is not particular to this problem space we address in this paper, but applicable to a range of problems in general, such as software~\cite{alasmary2019analyzing,KhormaliACNM20,AlasmaryAJAANM20} and network domains~\cite{chernikova2019fence}, where  the feature representation used for implementing the machine learning algorithm transforms the input by upholding a dependency among the features, which is not the case in the original image modality used in computer vision applications~\cite{abusnaina2021adversarial}.

\BfPara{Why Elevation and Actual Implications} We emphasize that we consider the elevation profile information as our inference input because this feature modality might be viewed, even to the most privacy-savvy individuals, as an innocent modality that does not necessarily expose much information about a specific location. Technically, however, it is not far-fetched to assume that there is an infinite number of mappings between a given elevation profile and location segments, which is shown to be not the case in this study under various plausible adversarial settings and objectives. 

Moreover, we recognize the subjectivity of valuing location privacy by users~\cite{fawaz2015anatomization,knijnenburg2013preference}. For instance, while some users might not care about sharing their exact location information all the time, some others may not feel comfortable sharing such information~\cite{krumm2009survey}. We demonstrate that such users, even when under the impression that they are not exposing their direct location, would be allowing an adversary to infer that location indirectly from the shared elevation. The sharing itself might put those users at risk, particularly when coupled with context. Imagine, for instance, an adversary who knows where a victim lives, but is able to infer that the activity of the victim, posted live, is associated with a location where the victim does not live, which would allow the adversary to stalk, or even break into the victim’s house. By the same token, an adversary who is able to precisely locate such an activity, in real-time, might be able to launch the physical attack possible under the sharing of the direct location (e.g., theft of expensive biking gear~\cite{Cybercasing}). 

\BfPara{Other Activities and Associated Risks} We note that, in general, mobility patterns are both individual and activity-dependent~\cite{hasan2013understanding}. For instance, the mobility of a salesperson would be totally different from the activity of a student. However, we note that our study is chiefly concerned with mobility patterns of specific activities: exercise. It is very difficult to envision a plausible scenario where a salesperson, for instance, would share the elevation profile of their activities moving door-to-door to sell a product. Similarly, while activities, in general, are tracked by various smartwatches (for everyday use), such tracking is limited to the high-level aggregate (e.g., total, sampled over epochs of time), and is not shareable directly. To this end, activities that are not exercise-related (on public trackable roads) are considered out of the scope of our use and attack models.

\BfPara{Explaining the Performance of Our Techniques} We note that, generally, the image and text representations outperformed the raw data used directly for our inference attack (except in the case of LSTM and random forest). The reason why LSTM performed well over the raw data is because of the natural mapping between such raw data (i.e., time series) and the operation of LSTM with attention to long sequences and their dependencies. In other words, LSTM is capable of representing the features of time series well. On the other hand, the random forest technique is known to be tolerant to noise, so one plausible explanation for its superior performance is its implicit feature selection and representation (by creating decision points on ranges). On the other hand, CNN worked well with image-like representations for its power in the representation of features from such modality. By the same token, SVM and MLP are shown to be superior in the text-like features for their tolerance to noise achieved by adjusting decision margins. 

In general, we also note that CONV1D performed better for \tma, but not for \tmb and \tmc, by demonstrating the power of CNN in general for capturing repeating patterns: \tma is concerned with profiling persons with a number of activities, some of which might be overlapping or even repeated, and CNN is known to extract high quality, representative, and discriminative features in that space, in contrast to the scenarios of \tmb and \tmc, where such patterns are less manifested.

Finally, we note that natural languages in general have various semantic and syntactic characteristics that are not manifested in the non-constrained domain of elevation profiles. To that end, while LSTM, when applied to features driven from the tf-idf or n-grams of a natural language utterance, would perform well, the same technique is not guaranteed to perform as well using the same feature representations, as we demonstrate in this study. One plausible explanation also for the superior performance of LSTM over the raw data is that the raw data explicitly and fully maintains an ordering that is essential for the operation of LSTM, whereas that ordering is implicit and for very short utterances in tf-idf and n-grams. Losing such valuable information as a result of feature extraction would (although marginally) affect the performance, as shown in this study.

\section{Related Work}\label{sec:related}
In this work, we addressed the problem of \emph{location privacy} in activity trackers using the side-channel information obtained from publicly shared elevation profiles. While there is no work that explicitly addresses this topic, there is has some studies on various topics that are related in the broad literature~\cite{narain2019security,narain2017perils,narain2019mitigating}.
In the following, we review some of those studies.

Most location privacy breaches are caused since users do not know why or how to preserve location privacy. 
\cite{aktypi2017a} developed a tool to examine possible privacy exposures of users in their social networks where the data is mostly collected from wearable devices. 
Using this tool, the authors aimed to enhance the awareness of information leakage in social networks, particularly fitness apps in which the data retrieved from wearable devices is shared on social networks.
Abdelmoty and Alrayes~\cite{Abdelmoty2017} aimed to increase awareness of location privacy on geo-social networks by surveying 186 users, where 77\% of them indicated they use location-based services often, several times a day, and 47\% of them reported that they were not aware that the location-based apps collect and store location information even when users select the private location option.
Moreover, 43\% of respondents were not aware that applications may share location information with third parties.

Despite the methods employed to preserve location privacy, several attacks are devised to uncover supposedly protected locations. 
Experiments for revealing exact locations from trajectories with private zones are conducted on a fitness-tracking social network, Strava~\cite{217618}. 
Researchers found the exact endpoints associated with users, even when such users selected the private zone option when sharing the training route. 
In another study, location trajectories of users are recovered from publicly available aggregated mobility data obtained from GSM operators~\cite{8356232}.
The attack relies on tracking the regularity---i.e., coming across the same location trace in the aggregated data regularly---and uniqueness---i.e., the location trace belongs to a unique user---of the user mobility traces to recover trajectories. 

As our study exemplifies, online social networks lay under the scope of privacy breach risks for users. 
Zheng \etal~\cite{DBLP:conf/icwsm/ZhengHYKL15} shows that sharing data that reveals spatiotemporal features of users' mobility patterns on online social networks reveals sensitive information such as home location, using a different form of data, \ie multimedia. 
Rossi \etal~\cite{DBLP:conf/icwsm/0004WSM15} show that location-based social networks are vulnerable to identity privacy breaches by revealing the identity of users by observing their mobility patterns.

Several attacks against general location privacy methods are proposed~\cite{Wernke2014}. 
The homogeneity attack~\cite{1617392} is an attack on k-anonymity to infer data of interest from other shared data. 
Machanavajjhala \etal~\cite{1617392} illustrated a scenario where an adversary infers the illness of a target person from available information, the zip code, age, etc. 
The same method can be applied to infer location data. 
In location distribution attacks~\cite{4417152}, the adversary exploits the fact that users are mostly not uniformly distributed in the location space.
Another attack by Shokri \etal~\cite{5958033} utilized the aggregated traffic statistics and environmental context information. 
The attack scenario includes an adversary who tries to reveal the possible location of the target by making use of the fact that the probability of the target's whereabouts is not uniformly distributed. 
Map matching methods \cite{10.1007/978-3-540-72037-9_8} aim to restrict the obfuscated area to a smaller but plausible area by removing irrelevant areas.
Movement boundary attacks were explored~\cite{Ghinita:2009:PVL:1653771.1653807}, where the adversary aims to calculate the movement boundary of a target by chasing the position queries and updates of the target. 
After calculating the boundary, the location of interest, such as home or workplace, is inferred and the irrelevant locations are discarded.

Although we did not directly touch upon preserving the location privacy in our study, there have been a few related studies in this space. 
The fast-growing need of preserving location privacy over the aforementioned attacks excited researchers' attention. 
Researchers introduce obfuscation methods such as decreasing the quality of the location by introducing inaccuracy and imprecision~\cite{Duckham:2005:FMO:2154273.2154286}. 
Additionally, the term k-anonymity is defined as obscuring the location information of individuals with \emph{k} number of other individuals within the region~\cite{Sweeney:2002:AKA:774544.774553,Gkoulalas-Divanis:2010:PKL:1882471.1882473}.

\section{Conclusion}\label{sec:conclusion}
In this paper, we presented new attacks on location privacy using only elevation profiles. 
The attacks are categorized into three types: predicting location by knowing the activity history of the target, predicting the borough by knowing the city of the target, and predicting the city of the target without any prior knowledge. 
The key contributions of our work are proving the concept that hiding the route of a workout and sharing only the elevation profile is not sufficient to preserve location privacy, defining a new attack surface by creating scenarios for possible threat models, and providing a machine-learning approach to realize such threat as attacks.
To validate our attacks we created three datasets by collecting data from athletes and mining data from a popular fitness-tracking website and Google Elevation API. 
We preprocessed the datasets by employing Natural Language Processing and Computer Vision approaches and then employed classification techniques to predict the location from elevation profiles. 
En route, we defined three threat models and evaluated each of them individually on the different datasets. 
As a result of the evaluations, we were able to identify the corresponding location of an elevation profile with accuracy between 59.59\% and 95.83\%. 

While this work highlights the clear trade-offs provided by the various defenses, their usability is largely unexplored. In our future work, we will explore the usability of compatible defenses such as devising and using route statistics that serves the same purpose as sharing elevation profile -- demonstrating the roughness of the route, while preserving users' privacy and acceptance.


\begin{IEEEbiography}[{\includegraphics[width=0.95in,height=1.25in,clip]{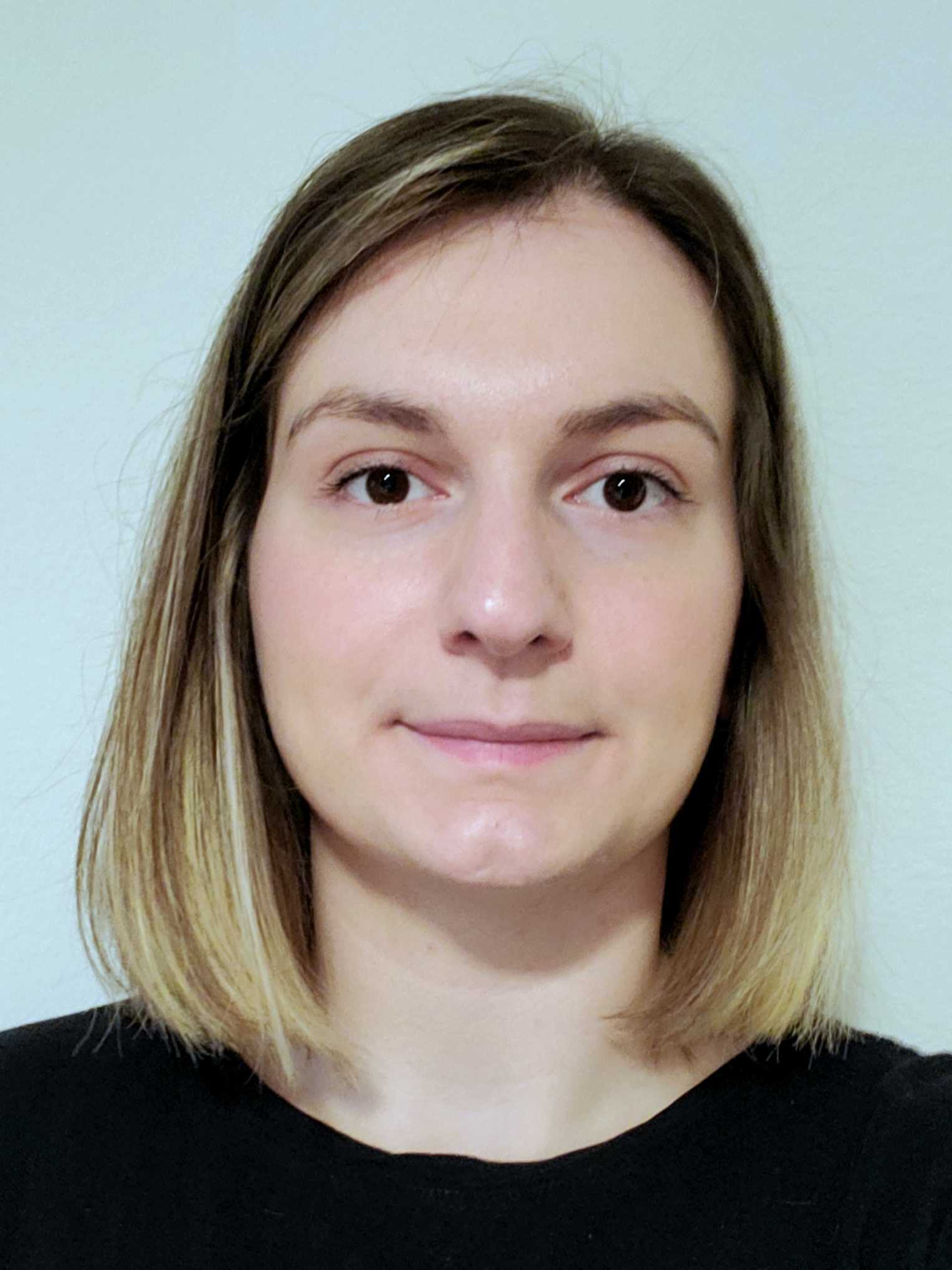}}]{Ülkü Meteriz-Yıldıran}
(S’20) obtained her Ph.D. and M.Sc. degrees from the University of Central Florida, in Computer Science with a Machine Learning concentration, in 2022 and 2022. She earned her B.Sc. degree in  Computer Engineering from Middle East Technical University in 2018. Her research interests are in applied machine learning on the privacy and security of wearable devices. She is a member of the Security and Analytics Lab (SEAL) at the University of Central Florida since 2018. In 2021, she interned at Amazon Web Services (AWS) as a software development engineer and is currently a machine learning researcher at Meta.
\end{IEEEbiography}

\begin{IEEEbiography}[{\includegraphics[width=0.95in,height=1.25in,clip]{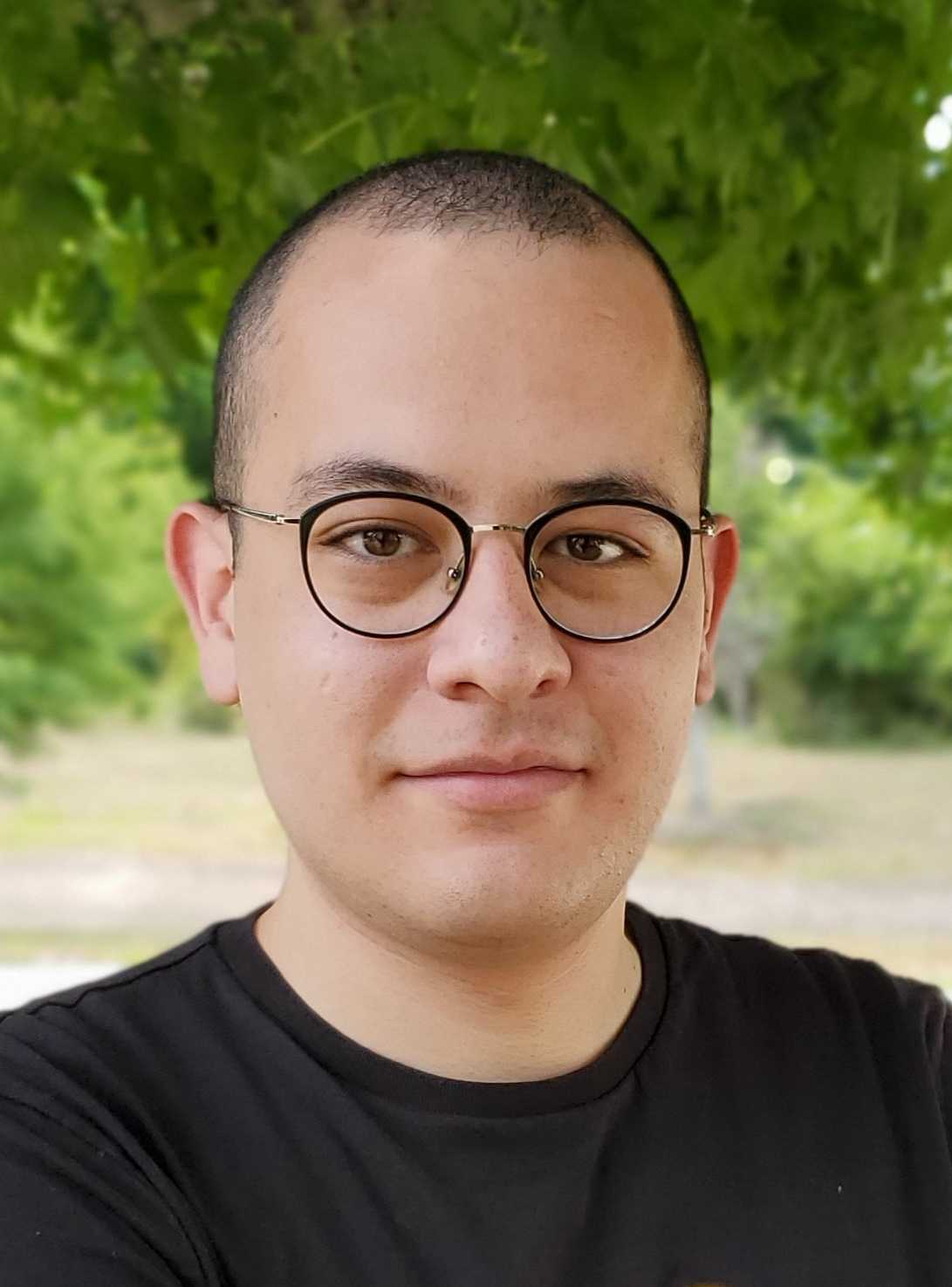}}]{Necip Fazıl Yıldıran} obtained his Ph.D. from the University of Central Florida in 2022 with a research focus on the analysis of highly configurable software. He earned his M.Sc. in Computer Science with a concentration on applied machine learning on the security and privacy of IoT, also from the University of Central Florida, in 2020. He earned his B.Sc. degree in Computer Engineering from Middle East Technical University in 2018. He interned at Google as a Software Engineer in 2020 and 2021, where he is currently a full-time Software Engineer.

\end{IEEEbiography}

\begin{IEEEbiography}[{\includegraphics[width=1in,height=1.25in,clip]{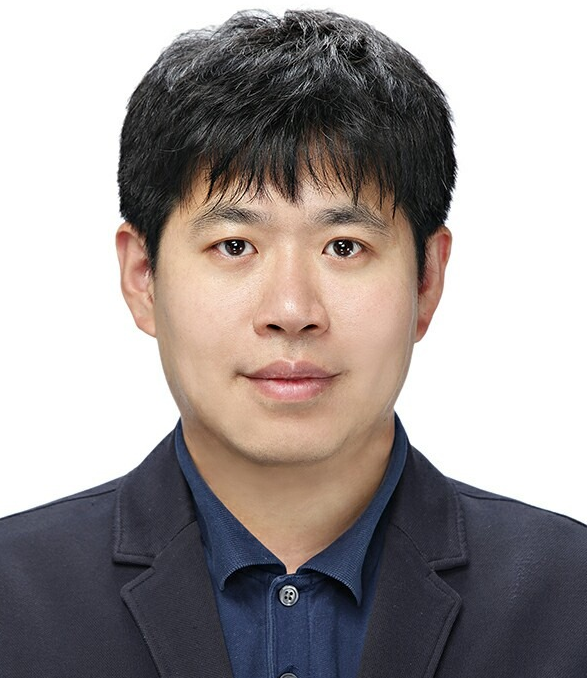}}]{Joongheon Kim}
(M'06--SM'18) has been with the School of Electrical Engineering, Korea University, Seoul, Korea, since 2019, where he is currently an associate professor. 
He received the B.S. and M.S. degrees in computer science and engineering from Korea University, Seoul, Korea, in 2004 and 2006, respectively; and the Ph.D. degree in computer science from the University of Southern California (USC), Los Angeles, CA, USA, in 2014. Before joining Korea University, he was with LG Electronics (Seoul, Korea, 2006--2009), InterDigital (San Diego, CA, USA, 2012), Intel Corporation (Santa Clara in Silicon Valley, CA, USA, 2013--2016), and Chung-Ang University (Seoul, Korea, 2016--2019). He serves as an associate editor for \textit{IEEE Transactions on Vehicular Technology}. He published more than 90 journals, 110 conference papers, and 6 book chapters. He also holds more than 50 granted patents\typeout{, majorly for 60\,GHz millimeter-wave IEEE 802.11ad, and IEEE 802.11ay standardization}. 

\end{IEEEbiography}

\begin{IEEEbiography}[{\includegraphics[width=1in,height=1.25in,clip,keepaspectratio]{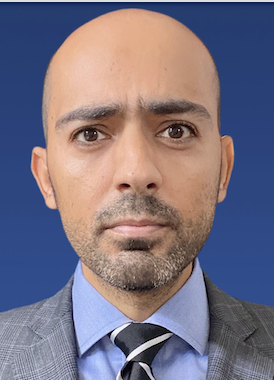}}]{David Mohaisen} (M'06-SM'15)
earned his M.Sc. and Ph.D. degrees from the University of Minnesota in 2011 and 2012, respectively. He is a Full Professor of Computer Science at the University of Central Florida, where he leads the Security and Analytics Lab (SEAL) and has been since 2017. Previously, he was an Assistant Professor at SUNY Buffalo (2015-2017) and a Senior Scientist at Verisign Labs (2012-2015). His research interests are in the broad area of applied security and privacy, covering aspects of computer and networked systems, software systems, IoT and AR/VR, and machine learning. His research has been supported by NSF, NRF, AFRL, AFOSR, etc., and has been published in top conferences and journals alike, with multiple best paper awards. His work was featured in the New Scientist, MIT Technology Review, ACM Tech News, Science Daily, etc. Among other services, he is currently an Associate Editor of IEEE Transactions on Mobile Computing and IEEE Transactions on Parallel and Distributed Systems.  He is a senior member of ACM (2018) and IEEE (2015).  
\end{IEEEbiography}\vspace{-10mm}

\end{document}